\documentclass[
  aps,
  pra,
  reprint,
  superscriptaddress,
  longbibliography]{revtex4-2}

\usepackage[T1]{fontenc}
\usepackage{amsmath,amssymb,bm}
\usepackage{mathrsfs}
\usepackage{booktabs}
\usepackage{graphicx}
\usepackage{tikz}
\usepackage{pgfplots}
\usepackage{microtype}
\usepackage{xcolor}
\usepackage{hyperref}
\usetikzlibrary{arrows.meta,positioning,calc}

\newcommand{\ket}[1]{\lvert #1\rangle}
\newcommand{\bc}{\boldsymbol{c}}
\newcommand{\Tr}{\operatorname{Tr}}
\newcommand{\vc}[1]{\vec{\boldsymbol{#1}}}

\allowdisplaybreaks[3]

\hypersetup{
  colorlinks = true,
  urlcolor = teal,
  citecolor = teal,
  linkcolor = teal
}

\pgfplotsset{compat=1.18}

\begin{document}

\title{Geometric construction of optimal probe states in distributed multiparameter quantum sensing}

\author{Hongzhen Chen}
\email{hzchen@szu.edu.cn}
\affiliation{Institute of Quantum Precision Measurement, State Key Laboratory of Radio Frequency Heterogeneous Integration, College of Physics and Optoelectronic Engineering, Shenzhen University, Shenzhen, China}

\author{Lingna Wang}
\email{lnwang@mae.cuhk.edu.hk}
\affiliation{Department of Mechanical and Automation Engineering, The Chinese University of Hong Kong, Shatin, Hong Kong SAR, China}

\author{Haidong Yuan}
\email{hdyuan@mae.cuhk.edu.hk}
\affiliation{Department of Mechanical and Automation Engineering, The Chinese University of Hong Kong, Shatin, Hong Kong SAR, China}

\date{\today}

\begin{abstract}
The choice of probe state is essential to achieving ultimate precision in quantum metrology. For a single parameter, an equal superposition of eigenstates associated with the largest and smallest generator eigenvalues is optimal. This simple construction does not directly extend to multiple parameters, since different parameters can favor different probes. We address this difficulty through a geometric construction of optimal probes for multiparameter estimation in distributed quantum sensing. In this setting, the generators commute and share a common eigenbasis, so each common eigenstate defines a vector of generator eigenvalues. We show that the minimal enclosing balls and ellipsoids of these vectors determine the maximal total quantum Fisher information and the minimum weighted covariance, respectively. In both cases, the boundary of the convex body identifies the common eigenstates that can contribute to an optimal probe, and its center and shape specify the conditions that their populations must satisfy. Our results provide a systematic approach to designing optimal probe states, linking the structure of the parameter generators directly to the attainable precision in multiparameter quantum metrology.
\end{abstract}

\maketitle

\section{Introduction}\label{sec:intro}

Quantum metrology uses quantum coherence and entanglement to improve the precision of estimating physical parameters~\cite{Helstrom1976Book,Holevo1982,BraunsteinCaves1994,GiovannettiLloydMaccone2011,DegenReinhardCappellaro2017,PezzeEtAl2018RMP}. Its applications include the estimation of multiple optical phases~\cite{HumphreysEtAl2013}, different components of a magnetic field~\cite{BaumgratzDatta2016}, and spatially distributed field properties~\cite{ZhangZhuang2021,BaamaraGessnerSinatra2023,FadelEtAl2023,LiEtAl2026Science,GuoLiuFadelHe2026}. These tasks often require several parameters to be estimated jointly using the same probe. For a fixed encoding and specified resources, the ultimate precision depends on the initial state, making the construction of optimal probes a fundamental problem in quantum parameter estimation~\cite{KaubrueggerEtAl2023,MeyerBorregaardEisert2021}.

For a single parameter encoded by unitary dynamics, the optimal probe has a simple spectral characterization. The quantum Fisher information (QFI) of a pure state equals four times the variance of the generator associated with the parameter~\cite{BraunsteinCaves1994,LiuYuanLuWang2020}. An equal superposition of eigenstates associated with the largest and smallest generator eigenvalues maximizes this variance~\cite{GiovannettiLloydMaccone2006}. The two spectral endpoints thus determine both an optimal initial state and the corresponding precision limit.

For multiple parameters, this simplicity is lost. Different generators can favor different probe states, so their separate single-parameter optima need not be attained by the same state~\cite{AlbarelliDemkowiczDobrzanski2022PRX}.
The spectral endpoints of each generator therefore no longer provide a complete prescription for the optimal probe state~\cite{PalEtAl2025}. Analytical constructions exist only for particular multiphase and multidimensional field estimation models~\cite{HumphreysEtAl2013,BaumgratzDatta2016, HouEtAl2020PRL}. While general semidefinite and conic formulations have also been developed for probe optimization~\cite{GoreckiEtAl2020,HayashiOuyang2024}, they are numerical in nature: they return an optimal probe for a given instance but offer little insight into how the generators determine its structure. A procedure that can determine the optimal probe state directly from the structure of the generators, comparable to the single-parameter prescription, is highly desirable.

In this work, we address this challenge in distributed quantum sensing. Distributed quantum sensing employs a network of spatially separated quantum sensors to measure a field or physical phenomenon that is distributed across multiple locations, with the goal of estimating global properties of the field rather than the local field values at individual sensors~\cite{ProctorKnottDunningham2018,GeEtAl2018,EldredgeEtAl2018,RubioEtAl2020,QianEtAl2021PRA,BringewattEtAl2021PRR,OhLeeLieJeong2020,OhJiangLee2022,MalitestaEtAl2023,EhrenbergBringewattGorshkov2023PRR,BringewattEtAl2024PRR,HongEtAl2025PRR,KimEtAl2025PRL,PezzeSmerzi2025PRL,GeJacobs2025PRL,GessnerSmerziPezze2020NatCommun,HamannSekatskiDuer2024,ZhuangZhangShapiro2018,TriggianiFacchiTamma2021}. In distributed quantum sensing, the local generators of distinct sensors act on separate Hilbert spaces and therefore commute. For any global parameter that is a linear combination of local parameters, its generator is the corresponding linear combination of local generators, thus also commute. These generators share a common eigenbasis, and each common eigenstate can be labeled by an eigenvalue vector whose entries are the eigenvalues of the generators associated with that state. We show that the structure of the optimal probe state is governed entirely by the geometry of these joint eigenvalue vectors of the generators: the minimum enclosing ball of these vectors determines the maximal total QFI (Sec.~\ref{sec:geometry}), while the optimal enclosing ellipsoid determines the minimum weighted covariance (Sec.~\ref{sec:weighted_probes}). In both cases, the optimal probe is a superposition of the extreme eigenvectors whose eigenvalue vectors lie on the boundary of the minimal enclosing ball/ellipsoid, with populations satisfying certain centroid conditions. This
is an exact multi-parameter generalization of the single-parameter prescription, to which it reduces when there is only one parameter. We illustrate the protocol with a simple model in Sec.~\ref{sec:two} and further apply it to field estimation in distributed quantum sensing in Sec.~\ref{sec:applications}.

\section{Multiparameter estimation in distributed quantum sensing}\label{sec:setting}
\subsection{quantum Cram\'er--Rao bound}
We first review the quantum Cram\'er--Rao bound (QCRB)~\cite{Helstrom1976Book,BraunsteinCaves1994}, which quantifies the local precision limit. 

For the estimation of a single parameter $\varphi$ encoded in a unitary evolution $U(\varphi)$, a probe initially prepared in $\rho_0$ evolves to $\rho_\varphi=U(\varphi)\rho_0U(\varphi)^\dagger$. For $\nu$ independent repetitions, the variance of a locally unbiased estimator is constrained by the quantum Cram\'er--Rao bound ,
\begin{equation}
	\operatorname{Var}(\widehat{\varphi})\geq\frac{1}{\nu F_Q},
\end{equation}
where $F_Q$ is the quantum Fisher information (QFI). For a pure input $\left|\psi_0\right\rangle$, the QFI is given by~\cite{LiuYuanLuWang2020},
\begin{equation}
	F_Q\left(\left|\psi_0\right\rangle\right)=4\left[\langle G_\varphi^2\rangle-\langle G_\varphi\rangle^2\right],
\end{equation}
Here
	\(G_\varphi=\left.iU^\dagger(\varphi)\frac{\partial U(\varphi)}{\partial\varphi}\right|_{\varphi=\varphi_0}\)
is the generator associated with the parameter, and the expectations are taken with respect to the initial probe state $|\psi_0\rangle$. Maximizing the QFI is therefore equivalent to maximizing the variance of the generator in the probe state.

 The variance of a Hermitian operator is maximized by placing equal populations on the eigenstates associated with the largest and smallest eigenvalues, $g_{\max}$ and $g_{\min}$. The optimal probe is thus~\cite{Audenaert2010}
\begin{equation}
	\left|\psi_0^{\mathrm{opt}}\right\rangle=\frac{\left|g_{\max}\right\rangle+e^{i\chi}\left|g_{\min}\right\rangle}{\sqrt{2}},
	\label{eq:single_optimal_probe}
\end{equation}
where the phase $\chi$ can be chosen arbitrarily. This state achieves $F_Q^{\max}=(g_{\max}-g_{\min})^2$, yielding
\begin{equation}
	\operatorname{Var}(\widehat{\varphi})\geq\frac{1}{\nu(g_{\max}-g_{\min})^2}.
\end{equation}
Geometrically, the optimal probe places equal populations at the two endpoints of the generator spectrum on the real line---the boundary points of the interval that contains all eigenvalues.

For multiple parameters, a probe that maximizes the QFI for one parameter need not maximize it for the others, even when the generators commute. Finding the optimal probe therefore requires accounting for the tradeoffs between the estimation errors of different parameters.

For the joint estimation of $\bm{\varphi}=(\varphi_1,\ldots,\varphi_m)^T$ under $U(\bm{\varphi})$, the generators are
\begin{equation}
	G_{\varphi_j}=\left.iU^\dagger(\bm{\varphi})\frac{\partial U(\bm{\varphi})}{\partial\varphi_j}\right|_{\bm{\varphi}=\bm{\varphi}_0}.
	\label{eq:generators}
\end{equation}
We assume that the generators are linearly independent modulo the identity.
In multiparameter estimation, the QFI is generalized to the quantum Fisher information matrix (QFIM)~\cite{SzczykulskaBaumgratzDatta2016}. For a pure input $\left|\psi_0\right\rangle$, it is~\cite{LiuYuanLuWang2020},
\begin{equation}
	\left[F_Q\left(\left|\psi_0\right\rangle\right)\right]_{jk}=4\operatorname{Re}\left[\left\langle G_{\varphi_j}G_{\varphi_k}\right\rangle-\left\langle G_{\varphi_j}\right\rangle\left\langle G_{\varphi_k}\right\rangle\right].
	\label{eq:generalqfi}
\end{equation}
A Positive Operator-Valued Measurement (POVM) $\mathsf M=\{M_y\}$ on the output gives outcome probabilities $p_y(\bm{\varphi})=\operatorname{Tr}(\rho_{\bm{\varphi}}M_y)$, with classical Fisher information matrix (CFIM)
\begin{equation}
	\left[F_C\right]_{jk}=\sum_{y:p_y(\bm{\varphi})>0}\frac{1}{p_y(\bm{\varphi})}\frac{\partial p_y(\bm{\varphi})}{\partial\varphi_j}\frac{\partial p_y(\bm{\varphi})}{\partial\varphi_k}.
	\label{eq:cfi}
\end{equation}
For $\nu$ independent repetitions, the covariance of any locally unbiased estimator is constrained by the classical and quantum Cram\'er--Rao bounds,
\begin{equation}
	\operatorname{Cov}\left(\widehat{\bm{\varphi}}\right)\succeq\frac{1}{\nu}F_C^{-1}\succeq\frac{1}{\nu}F_Q\left(\left|\psi_0\right\rangle\right)^{-1}.
	\label{eq:qcrb}
\end{equation}

In multiparameter estimation, the precision is typically quantified by a weighted covariance $\operatorname{Tr}\left[W\operatorname{Cov}\left(\widehat{\bm{\varphi}}\right)\right]$, where $W$ is real symmetric with $W\succ0$. The quantum Cram\'er--Rao bound gives
\begin{equation}
	\operatorname{Tr}\left[W\operatorname{Cov}\left(\widehat{\bm{\varphi}}\right)\right]\geq\frac{1}{\nu}\operatorname{Tr}\left[WF_Q\left(\left|\psi_0\right\rangle\right)^{-1}\right],
	\label{eq:cost}
\end{equation}
Our goal is to identify the probe states that minimize the attainable weighted covariance. Since the number of repetitions \(\nu\) contributes only a common factor of \(1/\nu\) to the cost, we omit it from the optimization. A singular QFIM leaves some parameter combinations unresolved and therefore has infinite cost for $W\succ0$. Singular directions and zero weights are discussed in Appendix~\ref{app:rank}.

\subsection{Distributed quantum sensing}
We consider a network of $M$ sensors, each acquiring a single phase $\theta_k(\bm{\varphi})$ through its interaction with the local field. The evolution of the network takes the form~\cite{GessnerSmerziPezze2020NatCommun}
\begin{equation}
	U(\bm{\varphi})=\bigotimes_{k=1}^{M}U^{(k)}\left(\theta_k(\bm{\varphi})\right).
\end{equation}
The generator associated with the local parameter $\theta_k$ is given by 
\begin{equation}
	G_{\theta_k}=\left.i\left(U^{(k)}(\theta_k)\right)^\dagger\frac{\partial U^{(k)}(\theta_k)}{\partial\theta_k}\right|_{\theta_k=\theta_{k,0}},
\end{equation}
where $\theta_{k,0}$ is the operating point. This generator only acts on the $k$-th sensor node, and Identity operators on the other sensors are implicit.
Applying the chain rule gives
\begin{equation}
	G_{\varphi_j}=\left.iU^\dagger(\bm{\varphi})\frac{\partial U(\bm{\varphi})}{\partial\varphi_j}\right|_{\bm{\varphi}=\bm{\varphi}_0}=\sum_{k=1}^{M}b_{jk}G_{\theta_k},
	\label{eq:network}
\end{equation}
where $b_{jk}=\left.\partial\theta_k/\partial\varphi_j\right|_{\bm{\varphi}_0}$.
Since $G_{\theta_k}$ acts only on sensor $k$, they commute with each other. Since $\{G_{\varphi_j}\}$ are linear combinations of commuting generators, therefore they commute as well.

\section{Joint spectrum and maximal total QFI}\label{sec:geometry}

Since the generators commute, we can choose a common eigenbasis $\{\left|i\right\rangle\}_{i=1}^d$ satisfying $G_{\varphi_j}\left|i\right\rangle=g_{j,i}\left|i\right\rangle$. Each common eigenstate thus corresponds to a point
\begin{equation}
	\vec{\boldsymbol{g}}_i=(g_{1,i},\ldots,g_{m,i})^T
\end{equation}
in an $m$-dimensional space. These points form the joint spectrum $\Lambda_{\mathrm{spec}}=\{\vec{\boldsymbol{g}}_i\}$. Any pure probe state can be expanded on $\{\left|i\right\rangle\}_{i=1}^d$ as
\begin{equation}
	\left|\psi_p\right\rangle=\sum_i\sqrt{p_i}\,e^{i\phi_i}\left|i\right\rangle,\qquad p_i\geq0,\quad\sum_i p_i=1,
	\label{eq:probe}
\end{equation}
which assigns a population $p_i$ to each spectral point. Coincident points can be combined because only their total population matters.

For this probe, the QFIM, which can be evaluated by Eq.~\eqref{eq:generalqfi}, is given by
\begin{equation}
	F_Q\left(\left|\psi_p\right\rangle\right)=4\sum_i p_i\left(\vec{\boldsymbol{g}}_i-\overline{\bm{g}}_p\right)\left(\vec{\boldsymbol{g}}_i-\overline{\bm{g}}_p\right)^T,
	\label{eq:covariance}
\end{equation}
where $\overline{\bm{g}}_p=\sum_i p_i\vec{\boldsymbol{g}}_i$ is the weighted centroid. The phases $\phi_i$ do not affect the QFIM.

\begin{figure}[t]
	\centering
	\includegraphics[width=\columnwidth]{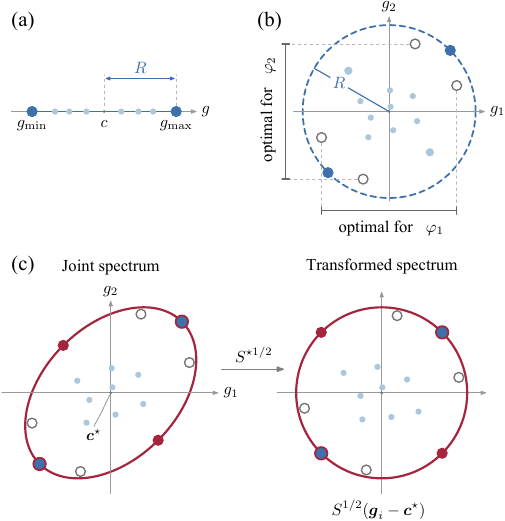}
	\caption{Geometry of optimal probes for commuting generators. (a) Single-parameter estimation. Equal populations at $g_{\min}$ and $g_{\max}$ maximize the QFI. (b) Maximal total QFI. Equal populations at the two blue contacts of the dashed minimum enclosing ball (MEB) maximize $\operatorname{Tr}F_Q$. Open circles mark the different pairs selected by separate estimation of $\varphi_1$ and $\varphi_2$. (c) Minimum weighted covariance ($W=I_2$). The red contacts of the optimal enclosing ellipse identify the eigenstates that can contribute to an optimal probe. The transformation $S^{1/2}(\bm g_i-\bm c^\star)$ maps the ellipse to a circle centered at the origin. Blue-filled contacts are also MEB contacts, and light-blue dots denote the remaining spectral points.}
	\label{fig:commuting_geometry}
\end{figure}

We first maximize the total QFI over all probe states. Taking the trace of Eq.~\eqref{eq:covariance} gives
\begin{equation}
	\operatorname{Tr}[F_Q(|\psi_p\rangle)]=4\sum_i p_i\|\vec{\boldsymbol{g}}_i-\overline{\bm g}_p\|^2.
	\label{eq:totalqfi}
\end{equation}
To bound this quantity, introduce an arbitrary reference point $\vec{\boldsymbol{c}}\in\mathbb R^m$. The variance decomposition gives
\begin{equation}
	\begin{aligned}
		\sum_i p_i\|\vec{\boldsymbol{g}}_i-\overline{\bm g}_p\|^2
		&=\sum_i p_i\|\vec{\boldsymbol{g}}_i-\vec{\boldsymbol{c}}\|^2-\|\overline{\bm g}_p-\vec{\boldsymbol{c}}\|^2\\
		&\leq\sum_i p_i\|\vec{\boldsymbol{g}}_i-\vec{\boldsymbol{c}}\|^2.
	\end{aligned}
	\label{eq:biasvariance}
\end{equation}
Since this holds for every $\vec{\boldsymbol{c}}$, choose the center $\vec{\boldsymbol{c}}^\star$ of the minimum enclosing ball (MEB) of the joint spectrum. Its radius $R$ satisfies $\|\vec{\boldsymbol{g}}_i-\vec{\boldsymbol{c}}^\star\|\leq R$ for every spectral point, so
\begin{equation}
	\operatorname{Tr}[F_Q(|\psi_p\rangle)]\leq4\sum_i p_i\|\vec{\boldsymbol{g}}_i-\vec{\boldsymbol{c}}^\star\|^2\leq4R^2.
	\label{eq:totalqfibound}
\end{equation}

The bound is attainable. To see this, we first establish that the center $\vec{\bc}^*$ lies in the convex hull of the \emph{contact set},  
\begin{equation}
I^* = \{ k : \|\vec{\boldsymbol{g}}_k - \vec{\bc}^*\| = R \},
\label{eq:contact-set}
\end{equation}
consisting of the eigenvalue vectors on the surface of the ball~\cite{ElzingaHearn1972,Gaertner1999,FischerGaertnerKutz2003,LimMcCann2022}. Suppose, for contradiction, that $\vec{\bc}^* \notin \mathrm{conv}(I^*)=\mathrm{conv}\{\vec{\boldsymbol{g}}_k|k\in I^*\}$. By the separating hyperplane theorem, there exists a vector $\vec{\boldsymbol{v}} \in \mathbb{R}^m$ and a scalar $\alpha$ such that
\begin{equation}
\vec{\boldsymbol{v}}^{\mathsf{T}} \vec{\boldsymbol{g}}_k > \alpha
\quad \text{for all } k \in I^*,
\qquad
\vec{\boldsymbol{v}}^{\mathsf{T}} \vec{\bc}^* < \alpha.
\end{equation}
Consider the perturbed point $\vec{\tilde{\bc}} = \vec{\bc}^* + \epsilon \vec{\boldsymbol{v}}$ for small $\epsilon > 0$. For $k \in I^*$, expanding the squared distance gives
\begin{equation}
\|\vec{\boldsymbol{g}}_k - \vec{\tilde{\bc}}\|^2
=
R^2 - 2\epsilon\, \vec{\boldsymbol{v}}^{\mathsf{T}}(\vec{\boldsymbol{g}}_k - \vec{\bc}^*) + O(\epsilon^2),
\end{equation}
and since $\vec{\boldsymbol{v}}^{\mathsf{T}}(\vec{\boldsymbol{g}}_k - \vec{\bc}^*) > 0$, the distance decreases for sufficiently small $\epsilon$. For $k \notin I^*$, we have $\|\vec{\boldsymbol{g}}_k - \vec{\bc}^*\| < R$ strictly, so by continuity the distance remains below $R$ for sufficiently small $\epsilon$. Hence $\vec{\tilde{\bc}}$ is the center of a ball of radius strictly smaller than $R$ that contains all eigenvalue vectors, contradicting the minimality of $R$. We thus have
\begin{equation}
\vec{\bc}^* \in \mathrm{conv}(I^*).
\label{eq:centroid-conv}
\end{equation}

This inclusion ensures the existence of a probability distribution $\{p_k^*\}$ supported exclusively on the contact set $I^*$ such that
\begin{equation}
\sum_{k \in I^*} p_k^* = 1,
\qquad
\sum_{k \in I^*} p_k^* \vec{\boldsymbol{g}}_k = \vec{\bc}^*.
\label{eq:centroid-condition}
\end{equation}
For the probe state with this distribution, the mean eigenvalue vector is $\bar{\boldsymbol{g}}^* = \vec{\bc}^*$, so the inequality in Eq.~\eqref{eq:biasvariance} is saturated, and
\begin{equation}
\Tr F_Q(|\psi_{\mathrm{opt}}\rangle)
= 4 \sum_{k \in I^*} p_k^* \|\vec{\boldsymbol{g}}_k - \vec{\bc}^*\|^2
= 4R^2.
\end{equation}
We thus establish the exact result
\begin{equation}
\max_{|\psi_0\rangle} \Tr F_Q(|\psi_0\rangle) = 4R^2,
\label{eq:max-total-QFI}
\end{equation}
which is achieved with the optimal probe state
\begin{equation}
|\psi_{\mathrm{opt}}\rangle
= \sum_{k \in I^*} \sqrt{p_k^*}\, e^{i\phi_k} |k\rangle,
\label{eq:optimal-probe-total}
\end{equation}
where $I^*$ is given by Eq.~\eqref{eq:contact-set} and $\{p_k^*\}$ satisfy the centroid condition in Eq.~\eqref{eq:centroid-condition}. The phases $\phi_k$ are arbitrary and do not affect the QFI.

This directly generalizes the single-parameter construction: the two extremal eigenvalues are replaced by the contact points of the minimum enclosing ball, and the equal populations become a distribution whose centroid coincides with the ball center.  Figure~\ref{fig:commuting_geometry}(a)  illustrates the single-parameter case, where equal populations at the two spectral endpoints give the largest generator variance. Figure~\ref{fig:commuting_geometry}(b) shows a two-parameter case, where maximizing the total QFI selects different pairs of eigenstates from maximizing the QFI for each parameter.

We next consider the weighted total QFI, defined for a positive semidefinite weight matrix $S \succeq 0$ as
\begin{equation}
\Tr[S F_Q(|\psi_0\rangle)]
= 4 \sum_{k=1}^d p_k
(\vec{\boldsymbol{g}}_k - \bar{\boldsymbol{g}})^{\mathsf{T}} S (\vec{\boldsymbol{g}}_k - \bar{\boldsymbol{g}}).
\label{eq:weighted-QFI}
\end{equation}
This figure of merit allows one to prioritize the QFI of certain parameters or combinations thereof, and it reduces to the total QFI when $S$ is the identity. Rewriting Eq.~\eqref{eq:weighted-QFI} as
\begin{equation}
\begin{aligned}
\Tr[S F_Q(|\psi_0\rangle)]
&= 4 \sum_{k=1}^d p_k
\|S^{1/2}\vec{\boldsymbol{g}}_k - S^{1/2}\bar{\boldsymbol{g}}\|^2,
\end{aligned}
\end{equation}
we see that the problem is equivalent to finding the minimal enclosing ball of the transformed points $\{S^{1/2}\vec{\boldsymbol{g}}_k\}$.
If the minimum enclosing ball of the transformed spectrum has squared radius
\begin{equation}
	R_S^2=\min_{\vec{\boldsymbol{c}}\in\mathbb{R}^m}\max_i\left(\vec{\boldsymbol{g}}_i-\vec{\boldsymbol{c}}\right)^T S\left(\vec{\boldsymbol{g}}_i-\vec{\boldsymbol{c}}\right),
	\label{eq:weightedradius}
\end{equation}
then
\begin{equation}
	\max_{\left|\psi_p\right\rangle}\operatorname{Tr}\left[SF_Q\left(\left|\psi_p\right\rangle\right)\right]=4R_S^2.
	\label{eq:weightedmeb}
\end{equation}
The optimal probe takes the form $|\psi_{\mathrm{opt}}\rangle
= \sum_{k \in I^*} \sqrt{p_k^*}\, e^{i\phi_k} |k\rangle$, with the contact set now defined with respect to the transformed metric,
\begin{equation}
I^* = \left\{ k : \|S^{1/2}\vec{\boldsymbol{g}}_k - S^{1/2}\vec{\bc}_S^*\| = R_S \right\},
\label{eq:contact-set-weighted}
\end{equation}
or equivalently,
\begin{equation}
I^* = \left\{ k : (\vec{\boldsymbol{g}}_k - \vec{\bc}_S^*)^{\mathsf{T}} S (\vec{\boldsymbol{g}}_k - \vec{\bc}_S^*) = R_S^2 \right\}.
\end{equation}
The populations satisfy the centroid condition in the transformed coordinates,
\begin{equation}
\sum_{k \in I^*} p_k^* = 1,
\qquad
S^{1/2}\!\left( \sum_{k \in I^*} p_k^* \vec{\boldsymbol{g}}_k - \bc_S^* \right) = 0.
\end{equation} 
Geometrically, the weighted total QFI is characterized by the \emph{minimum enclosing ellipsoid} of all eigenvalue vectors, 
\begin{equation}
(\vec{\boldsymbol{g}}_k - \vec{\bc}_S^*)^{\mathsf{T}} S (\vec{\boldsymbol{g}}_k - \vec{\bc}_S^*) \leq R_S^2,
\end{equation}
whose shape and orientation are determined by $S$.
As shown in  Fig.~\ref{fig:commuting_geometry}(c), the centered transformation $S^{1/2}(\vec{\boldsymbol{g}}_i-\vec{\boldsymbol{c}}^\star)$ maps the optimal ellipse to a circle centered at the origin with the contacts remaining on the boundary and the interior points remaining inside.

\section{Minimum weighted covariance and optimal probes}\label{sec:weighted_probes}

\subsection{Minimum weighted covariance}\label{sec:precision}

We now turn from the total or weighted QFI to the weighted covariance $\mathcal C_W=\operatorname{Tr}[W\operatorname{Cov}(\widehat{\bm{\varphi}})]$, where $W\succ0$ specifies the relative importance of the estimation errors. From the QCRB, we have  \begin{equation}
    \mathcal C_W\geq \operatorname{Tr}[WF_Q(|\psi_p\rangle)^{-1}].
\end{equation} For commuting generators, the weak-commutativity condition $\langle[G_{\varphi_j},G_{\varphi_k}]\rangle=0$ holds for any $|\psi_0\rangle$. The QCRB is therefore locally attainable~\cite{Matsumoto2002,RagyJarzynaDemkowiczDobrzanski2016,PezzeEtAl2017PRL,YangPangZhouJordan2019}, and the minimum attainable weighted covariance is thus
\begin{equation}
	C_W^\star=\min_{|\psi_p\rangle}\operatorname{Tr}\left[WF_Q(|\psi_p\rangle)^{-1}\right].
	\label{eq:cstar}
\end{equation}
Since for singular QFIMs, $C_W\rightarrow \infty$, the minimization is thus always achieved at a probe state that has nonsingular QFIM.

This cost can be related to the enclosing geometry via the matrix Cauchy--Schwarz inequality. For $F,W\succ0$ and $S\succeq0$, we have
\begin{equation}
	\operatorname{Tr}(WF^{-1})\operatorname{Tr}(SF)\geq\left(\operatorname{Tr}\sqrt{S^{\frac12}WS^{\frac12}}\right)^2.
	\label{eq:matrixcs}
\end{equation}
For nonzero $S$, equality holds when $S=\gamma F^{-1}WF^{-1}$ for any $\gamma>0$, or equivalently
\begin{equation}
	F=\sqrt\gamma\,W^{1/2}\left(W^{1/2}SW^{1/2}\right)^{-1/2}W^{1/2}.
	\label{eq:csqfim}
\end{equation}
The inequality and its equality condition are proved in Appendix~\ref{app:matrixcs}. Combining it with $\operatorname{Tr}[SF_Q(|\psi_p\rangle)]\leq4R_S^2$ gives
\begin{equation}
	\begin{aligned}
		&\operatorname{Tr}\left[WF_Q(|\psi_p\rangle)^{-1}\right]\\
		&\quad\geq\frac{\left(\operatorname{Tr}\sqrt{S^{\frac12}WS^{\frac12}}\right)^2}{\operatorname{Tr}[SF_Q(|\psi_p\rangle)]}
		\geq\frac{\left(\operatorname{Tr}\sqrt{S^{\frac12}WS^{\frac12}}\right)^2}{4R_S^2}.
	\end{aligned}
	\label{eq:Sbound}
\end{equation}
This holds for every pure probe with a nonsingular QFIM and every nonzero $S\succeq0$. Maximizing over $S$ therefore yields
\begin{equation}
	C_W^\star\geq\max_{S\succeq0,\,S\ne0}\frac{\left(\operatorname{Tr}\sqrt{S^{\frac12}WS^{\frac12}}\right)^2}{4R_S^2}.
	\label{eq:geometriclowerbound}
\end{equation}

This bound is also tight. To show it, let \(|\psi_{p^\star}\rangle=\sum_k \sqrt{p_k^\star}e^{i\phi_k}|k\rangle\) be a minimizing probe of $C_W=\Tr[WF_Q^{-1}(|\psi_{p}\rangle)]$ with $F_Q^\star=F_Q(|\psi_{p^\star}\rangle)\succ0$, and choose
\begin{equation}
	S^\star=(F_Q^\star)^{-1}W(F_Q^\star)^{-1}.
	\label{eq:sstar}
\end{equation}
This choice saturates the matrix Cauchy--Schwarz inequality. To show that the same probe also saturates $\operatorname{Tr}[S^{\star}F_Q(|\psi_p\rangle)]\leq4R_{S^\star}^2$, consider an arbitrary probe state \(|\psi_{q}\rangle\) with populations
\(\{q_k\}\), and define the interpolating population distribution $p_t=(1-t)p^\star+tq$, with $0\leq t\leq1$. The corresponding probe state is denoted by
\begin{equation}
    |\psi_{p_t}\rangle
    :=
    \sum_k\sqrt{p_k(t)}\,e^{i\phi_k(t)}\ket{k},
    \label{eq:interpolating-state}
\end{equation}
where the phases may be chosen arbitrarily. Let $\bm d=\overline{\bm g}_{p^\star}-\overline{\bm g}_q$, where $\bar{\boldsymbol{g}}_{p^\star}=
    \sum_k p_k^\star\vec{\boldsymbol{g}}_k$, $\bar{\boldsymbol{g}}_{q}=
    \sum_k q_k\vec{\boldsymbol{g}}_k$.
A direct computation of $F_Q(|\psi_{p_t}\rangle)$ yields
\begin{equation}
	\begin{aligned}
		F_Q(|\psi_{p_t}\rangle)={}&(1-t)F_Q^\star+tF_Q(|\psi_q\rangle)\\
		&+4t(1-t)\bm d\bm d^T.
	\end{aligned}
	\label{eq:populationpath}
\end{equation}
Since $p^\star$ is optimal, $\operatorname{Tr}\left[WF_Q(|\psi_{p_t}\rangle)^{-1}\right]$ cannot decrease as $t$ increases from zero. We thus have
\begin{equation}
	\begin{aligned}
		0&\leq\left.\frac{d}{dt}\operatorname{Tr}\left[WF_Q(|\psi_{p_t}\rangle)^{-1}\right]\right|_{t=0^+}\\
		&=\operatorname{Tr}(S^\star F_Q^\star)-\operatorname{Tr}[S^\star F_Q(|\psi_q\rangle)]-4\bm d^TS^\star\bm d,
	\end{aligned}
	\label{eq:populationoptimality}
\end{equation}
where we have used the fact that \(
\frac{d}{dt}F^{-1}
=
-F^{-1}\frac{dF}{dt}F^{-1}.
\)
Because $\bm d^TS^\star\bm d\geq0$, this implies $\operatorname{Tr}[S^\star F_Q(|\psi_q\rangle)]\leq\operatorname{Tr}(S^\star F_Q^\star)$ for every $q$. Thus the optimal probe also maximizes the weighted total QFI for $S^\star$, giving $\operatorname{Tr}(S^\star F_Q^\star)=4R_{S^\star}^2$ by Eq.~\eqref{eq:weightedmeb}. The same probe therefore saturates both inequalities in Eq.~\eqref{eq:Sbound}. Moreover, $4R_{S^\star}^2=\operatorname{Tr}(S^\star F_Q^\star)=\operatorname{Tr}[W(F_Q^\star)^{-1}]=C_W^\star$, and  $\operatorname{Tr}\sqrt{S^{\star\frac12}WS^{\star\frac12}}=C_W^\star$. We thus have $\frac{\left(\operatorname{Tr}\sqrt{S^{\star\frac12}WS^{\star\frac12}}\right)^2}{4R_{S^\star}^ 2}=C_W^\star$, which shows that the bound is indeed achievable.
More details of the proof can be found in Appendix~\ref{app:populationvariation}.
In Appendix~\ref{app:sdp}, we further show that the bound can be computed via a semidefinite programming (SDP),
\begin{equation}
	\begin{aligned}
		\operatorname*{maximize}_{S,X,\bm q,t}\;&\operatorname{Tr}X\\
		\text{subject to }&\begin{pmatrix}S&X\\X^T&W\end{pmatrix}\succeq0,\\
		&\begin{pmatrix}S&\bm q\\\bm q^T&t\end{pmatrix}\succeq0,\\
		&\vec{\boldsymbol{g}}_i^TS\vec{\boldsymbol{g}}_i-2\bm q^T\vec{\boldsymbol{g}}_i+t\leq1\quad\text{for all }i.
	\end{aligned}
	\label{eq:ellipsoid_sdp}
\end{equation}
Here $X$, $\bm q$, $t$ are auxiliary variables with $X$ a real $m\times m$ matrix, $\bm q\in\mathbb R^m$, and $t\in\mathbb R$. The center of the optimal ellipsoid can be obtained from the output of the SDP as $\vec{\boldsymbol{c}}^\star=(S^\star)^{-1}\bm q^\star$, the optimal cost can be obtained as $C_W^\star=(\operatorname{Tr}X^\star)^2/4$ and the optimal QFIM is given by 
\begin{equation}
	F_Q^\star=\frac{4R_{S^\star}^2}{\operatorname{Tr}\sqrt{S^{\star\frac12}WS^{\star\frac12}}}W^{1/2}\left(W^{1/2}S^\star W^{1/2}\right)^{-1/2}W^{1/2}.
	\label{eq:targetqfim}
\end{equation}

The optimal probe that saturates the bound takes the form $|\psi_{\mathrm{opt}}\rangle
= \sum_{k \in I^*} \sqrt{p_k^*}\, e^{i\phi_k} |k\rangle$ with the contact set
\begin{equation}
I^* = \left\{ k : \|S^{\star 1/2}\vec{\boldsymbol{g}}_k - S^{\star1/2}\vec{\bc}^*\| = R_{S^\star} \right\},
\label{eq:contact-set-variance}
\end{equation}
and the populations $\{p_k\}$ satisfy
\begin{equation}
	\begin{gathered}
		\sum_{k\in I^\star}p_k=1,\qquad S^{\star1/2}\left(\sum_{k\in I^\star}p_k\vec{\boldsymbol{g}}_k-\vec{\boldsymbol{c}}^\star\right)=0,\\
		4\sum_{k\in I^\star}p_k(\vec{\boldsymbol{g}}_k-\vec{\boldsymbol{c}}^\star)(\vec{\boldsymbol{g}}_k-\vec{\boldsymbol{c}}^\star)^T=F_Q^\star.
	\end{gathered}
	\label{eq:momentmatching}
\end{equation}

In summary, given the commuting generators $G_{\varphi_1},\ldots,G_{\varphi_m}$ and the weight matrix $W\succ0$, an optimal probe can be constructed in three steps:

(i) For each common eigenstate $\ket{i}$, find $g_{j,i}$ such that $G_{\varphi_j}\ket{i}=g_{j,i}\ket{i}$, and form the joint eigenvalue vectors $\vec{\boldsymbol{g}}_i=(g_{1,i},\ldots,g_{m,i})^T$.

(ii) Determine the optimal enclosing ellipsoid for the given $W$ and identify its contact points. Read $S^\star$ and squared radius $R_{S^\star}^2$ from the ellipsoid equation, and obtain the target QFIM from Eq.~\eqref{eq:targetqfim}.

(iii) Solve Eq.~\eqref{eq:momentmatching} for nonnegative populations $\{p_i\}$ on these contact points. Each solution gives an optimal probe
\begin{equation}
	\ket{\psi_p}=\sum_{i\in I^\star}\sqrt{p_i}\,e^{i\phi_i}\ket{i},
\end{equation}
with arbitrary phases $\phi_i$.

\subsection{Two-parameter geometry}\label{sec:two}

For two parameters, the precision bound has a nice geometric interpretation in terms of the semiaxes of an enclosing ellipse. We first introduce the transformed coordinates $\widetilde{\vec{\boldsymbol{g}}}_i=W^{-1/2}\vec{\boldsymbol{g}}_i$, $\widetilde{\vec{\boldsymbol{c}}}=W^{-1/2}\vec{\boldsymbol{c}}$, $\widetilde S=W^{1/2}SW^{1/2}$. The quadratic form is invariant under this change with $(\vec{\boldsymbol{g}}_k-\vec{\boldsymbol{c}})^\top S(\vec{\boldsymbol{g}}_k-\vec{\boldsymbol{c}})=(\tilde{\boldsymbol{g}}_k-\tilde{\boldsymbol{c}})^\top\widetilde{S}(\tilde{\boldsymbol{g}}_k-\tilde{\boldsymbol{c}})$. The enclosing radius is therefore the same in the two coordinate systems. Additionally, in these transformed coordinates, we have $\widetilde{F_Q}=W^{-1/2}F_QW^{-1/2}$ and Eq.(\ref{eq:targetqfim}) can be equivalently written as 
\begin{equation}\label{eq:transformedQFIM}
    \widetilde{F_Q^\star}=\frac{4R_{\widetilde S^\star}^2}{\Tr\sqrt{\widetilde S^\star}}\widetilde S^{\star -\frac12}.
\end{equation}

Let $\lambda_1,\lambda_2>0$ be the eigenvalues of $\widetilde S$. The semiaxes of the enclosing ellipse, $(\tilde{\boldsymbol{g}}_k-\tilde{\boldsymbol{c}})^\top\widetilde{S}(\tilde{\boldsymbol{g}}_k-\tilde{\boldsymbol{c}})=R_S^2$, are given by $a=R_S/\sqrt{\lambda_1}$ and $b=R_S/\sqrt{\lambda_2}$. Since $S^{1/2}WS^{1/2}$ and $W^{1/2}SW^{1/2}$ have the same eigenvalues, we have $\operatorname{Tr}\sqrt{S^{\frac12}WS^{\frac12}}=\operatorname{Tr}\sqrt{ \widetilde S}=\sqrt{\lambda_1}+\sqrt{\lambda_2}$, thus
\begin{equation}
	\begin{aligned}
		\frac{\left(\operatorname{Tr}\sqrt{S^{\frac12}WS^{\frac12}}\right)^2}{4R_S^2}&=\frac{(\sqrt{\lambda_1}+\sqrt{\lambda_2})^2}{4R_S^2}\\
		&=\frac14\left(\frac1a+\frac1b\right)^2=\frac1{h^2},
	\end{aligned}
\end{equation}
where $h=2ab/(a+b)$ is the harmonic mean of the semiaxes. Thus
\begin{equation}
	C_W^\star=\frac{1}{h_\star^2},\quad \text{with}\ h_\star=\min_{\mathcal E\supset\widetilde\Lambda_{\mathrm{spec}}}\frac{2a(\mathcal E)b(\mathcal E)}{a(\mathcal E)+b(\mathcal E)},
	\label{eq:harmonic}
\end{equation}
where $\widetilde\Lambda_{\mathrm{spec}}=\{W^{-1/2}\vec{\boldsymbol{g}}_i\}$ and $\mathcal E\supset\widetilde\Lambda_{\mathrm{spec}}$ denotes ellipsoids enclosing $\widetilde\Lambda_{\mathrm{spec}}$.

When the convex hull of the eigenvalue vectors is a triangle, the optimal enclosing ellipse and probe have a nice geometrical construction. Suppose the convex hull of \(\widetilde{\vec{g}} = W^{-1/2}\vec{g}\) is a triangle with area \(A\), perimeter \(L\), and inradius \(r = 2A/L\). In Appendix~\ref{app:triangle}, we show that the unique optimal ellipse is centered at the triangle's incenter, with each vertex lying on its boundary, and it has the harmonic mean \(h^\star = 2r\). The optimal probe is supported at the three vertices, with populations proportional to the lengths of their opposite sides, \(p_i^\star = \ell_i/L\). This yields the optimal weighted covariance
\begin{equation}
  C_W^\star =\frac{1}{h^{\star 2}}=\frac{1}{4r^2}= \frac{L^2}{16A^2}.  
\end{equation}

\begin{figure}[t]
	\centering
	\includegraphics[width=0.46\textwidth]{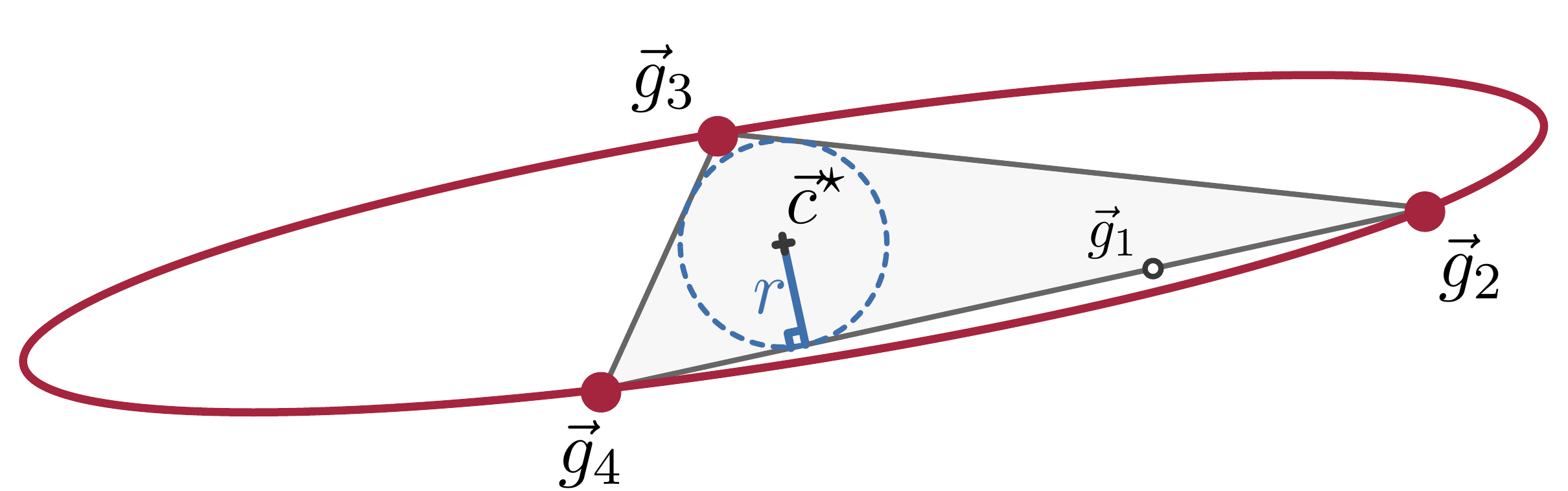}
	\caption{Geometry of the toy model for $W=I_2$. The optimal enclosing ellipse (solid red) passes through $\vec{\boldsymbol{g}}_2,\vec{\boldsymbol{g}}_3,\vec{\boldsymbol{g}}_4$ and shares its center $\vec{\boldsymbol{c}}^\star$ with the triangle's incircle (dashed blue).}
	\label{fig:toy_geometry}
\end{figure}

For example, consider a toy model with two commuting generators 
\begin{equation}
	G_{\varphi_1}=\begin{pmatrix}
		1&0&0&0\\
		0&2&0&0\\
		0&0&-2&0\\
		0&0&0&-1
	\end{pmatrix},\quad
	G_{\varphi_2}=\begin{pmatrix}
		1&0&0&0\\
		0&3&0&0\\
		0&0&-1&0\\
		0&0&0&-3
	\end{pmatrix}.
	\label{eq:toy_g}
\end{equation}
The common eigenstates are  $\left|k\right\rangle$, $k=1,\ldots,4$ with the associated eigenvalue vectors given by $\vec{\boldsymbol{g}}_1=(1,1)^T$, $\vec{\boldsymbol{g}}_2=(2,3)^T$, $\vec{\boldsymbol{g}}_3=(-2,-1)^T$, and $\vec{\boldsymbol{g}}_4=(-1,-3)^T$. Since $\vec{\boldsymbol{g}}_1=(2\vec{\boldsymbol{g}}_2+\vec{\boldsymbol{g}}_4)/3$, their convex hull is the triangle formed by $\vec{\boldsymbol{g}}_2,\vec{\boldsymbol{g}}_3,\vec{\boldsymbol{g}}_4$. For $W=I_2$, the optimal enclosing ellipse is centered at the triangle's incenter $\vec{\boldsymbol{c}}^\star=(-1,2-\sqrt{10})^T$ and passes through $\vec{\boldsymbol{g}}_2,\vec{\boldsymbol{g}}_3,\vec{\boldsymbol{g}}_4$, as shown in Fig.~\ref{fig:toy_geometry}. 
The minimum harmonic mean of the semiaxes is
\begin{equation}
	h^\star=\frac{4A}{L}=2(\sqrt5-\sqrt2).
	\label{eq:toy_harmonic}
\end{equation}

The contact points are $\vec{\boldsymbol{g}}_2,\vec{\boldsymbol{g}}_3,\vec{\boldsymbol{g}}_4$, while $\vec{\boldsymbol{g}}_1$ lies strictly inside the ellipse. The optimal probe therefore has support on $\left|2\right\rangle$, $\left|3\right\rangle$, and $\left|4\right\rangle$ with
\begin{equation}
	(p_2^\star,p_3^\star,p_4^\star)=\frac{(\sqrt5,3\sqrt5,4\sqrt2)}{4(\sqrt5+\sqrt2)}.
	\label{eq:toy_populations}
\end{equation}
An optimal probe is therefore
\begin{equation}
	\left|\psi^\star\right\rangle=\sum_{k=2}^4\sqrt{p_k^\star}\left|k\right\rangle.
	\label{eq:toy_probe}
\end{equation}
The relative phases of the amplitudes can be chosen arbitrarily. The minimum cost is
\begin{equation}
	C_{I_2}^\star=\frac{1}{h^{* 2}}=\frac{7+2\sqrt{10}}{36}.
	\label{eq:toy_cost}
\end{equation}
Detailed derivations can be found in  Appendix~\ref{app:triangle}.

When the convex hull of \(\widetilde{\vec{g}} = W^{-1/2}\vec{g}\) is a parallelogram, we show in Appendix~\ref{app:parallelogram} that $C_W^\star=\frac{(d_1+d_2)^2}{A^2}$, where $d_1$, $d_2$ are the distances from the center of the parallelogram to two adjacent vertices,
and $A$ is the area of the parallelogram.

\section{Applications to distributed quantum sensing}\label{sec:applications}

We provide two examples to demonstrate the general protocol. The first example considers the joint estimation of the mean and gradient of a field, demonstrating how the covariance weight changes the optimal probe. The second example considers a localized-source model, illustrating an analytic construction for a larger sensor network.

\subsection{Joint estimation of a field mean and gradient}\label{sec:pairmotivation}

Consider a two-qubit network for estimating a field mean $\varphi_m$ and gradient $\varphi_g$~\cite{ApellanizEtAl2018,AltenburgEtAl2017}. The qubits are at dimensionless positions $x_1=-1$ and $x_2=1$, with local phases $\theta_k=\varphi_m+x_k\varphi_g$ and $U^{(k)}(\theta_k)=e^{-i\theta_k\sigma_z^{(k)}}$. Using $\sigma_z\left|0\right\rangle=\left|0\right\rangle$ and $\sigma_z\left|1\right\rangle=-\left|1\right\rangle$, the generators are
\begin{equation}
	G_{\varphi_m}=\sigma_z^{(1)}+\sigma_z^{(2)},\qquad G_{\varphi_g}=-\sigma_z^{(1)}+\sigma_z^{(2)}.
	\label{eq:pairgen}
\end{equation}

\begin{figure}[t]
	\centering
	\includegraphics[width=\columnwidth]{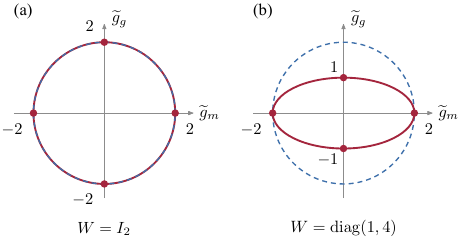}
	\caption{Effect of the covariance weight in the two-node field--gradient model. Points show the transformed joint spectrum $(\widetilde g_m,\widetilde g_g)^T=W^{-1/2}\vec{\boldsymbol{g}}_i$ for (a) $W=I_2$ and (b) $W=\operatorname{diag}(1,4)$. Solid red curves are optimal enclosing ellipses, and dashed blue curves are minimum enclosing circles. In (a), the two curves coincide and contact all four points. In (b), the circle contacts only the horizontal pair, while the ellipse contacts all four points.}
	\label{fig:weights}
\end{figure}

The joint spectrum consists of the four points $\{\vec{\boldsymbol{g}}_1=(2,0)^T,\vec{\boldsymbol{g}}_2=(0,-2)^T,\vec{\boldsymbol{g}}_3=(0,2)^T,\vec{\boldsymbol{g}}_4=(-2,0)^T\}$ shown in Fig.~\ref{fig:weights}(a), with the associated eigenstates given by $|00\rangle$, $|01\rangle$, $|10\rangle$ and $|11\rangle$ respectively. The minimal enclosing ball is the circle centered at the origin with radius $R_S=2$. The maximal total QFI is thus given by $\max \Tr(F_Q)=4R_S^2=16$. Since all four spectral points lie on the boundary of the minimal enclosing ball, the optimal probe that maximizes the total QFI takes the form $|\psi_{\rm opt}\rangle=\sqrt{p_{00}}e^{i\phi_1}|00\rangle+\sqrt{p_{01}}e^{i\phi_2}|01\rangle+\sqrt{p_{10}}e^{i\phi_3}|10\rangle+\sqrt{p_{11}}e^{i\phi_4}|11\rangle$ with $p_{00}=p_{11}=p$ and $p_{10}=p_{01}=\frac12-p$ determined from the centroid condition.
Indeed, the QFIM of such states is
\begin{equation}
	F_Q(|\psi(p)\rangle)=\begin{pmatrix}32p&0\\0&16-32p\end{pmatrix},
	\label{eq:pairqfim}
\end{equation}
which achieves the maximal total QFI.

For the weighted covariance with  $W=\operatorname{diag}(w_m,w_g)\succ0$, the transformed spectral points $W^{-1/2}\vec{\boldsymbol{g}}_i$ are $(\pm 2/\sqrt{w_m},0)^T$ and $(0,\pm 2/\sqrt{w_g})^T$.
The optimal enclosing ellipse is the ellipse that is centered at the origin with the semiaxes given by $a=2/\sqrt{w_m}$ and $b=2/\sqrt{w_g}$. This ellipse can be written as $\widetilde{\vec{\boldsymbol{g}}}_k^{\mathsf{T}} \widetilde S^\star \widetilde{\vec{\boldsymbol{g}}}_k = R_{S^{\star}}^2$ with $\widetilde S^\star=\operatorname{diag}(w_m,w_g)$ and $R_{S^\star}=2$. The minimal weighted covariance can be obtained as
\begin{equation}
	C_W^\star=\frac14\left(\frac1a+\frac1b\right)^2=\frac{(\sqrt{w_m}+\sqrt{w_g})^2}{16}.
\end{equation}
Fig.~\ref{fig:weights} shows the change of the minimal enclosing ellipse from $W=I_2$ in Fig.~\ref{fig:weights}(a) to $W=\operatorname{diag}(1,4)$ in panel (b).
 
In the original coordinates, $S^\star=W^{-1/2}\widetilde S^\star W^{-1/2}=I_2$. Equation~\eqref{eq:targetqfim} gives
\begin{equation}
	F_Q^\star=\frac{16}{\sqrt{w_m}+\sqrt{w_g}}W^{1/2}.
	\label{eq:pairtargetqfim}
\end{equation}
Matching Eq.~\eqref{eq:pairqfim} to this target fixes the populations,
\begin{equation}
	p^\star=\frac{\sqrt{w_m}}{2(\sqrt{w_m}+\sqrt{w_g})}.
	\label{eq:pairresult}
\end{equation}
The optimal probe state thus has $p_{00}^\star=p_{11}^\star=p^\star$ and $p_{01}^\star=p_{10}^\star=\tfrac12-p^\star$.

\subsection{Distributed sensing of a localized source}\label{sec:source_application}

\begin{figure}[t]
	\centering
	\includegraphics[width=0.8\columnwidth]{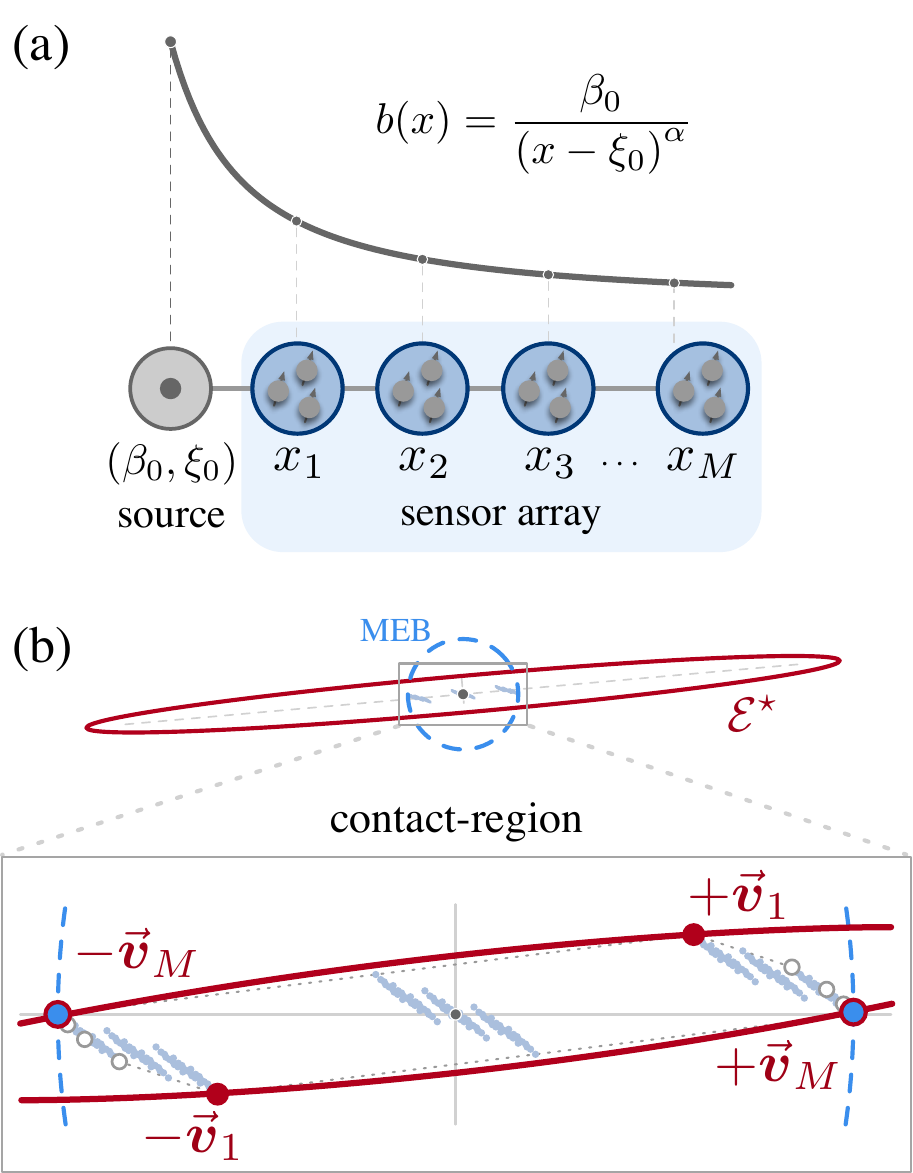}
	\caption{Localized-source sensing and construction of optimal probes. (a) Schematic of an array of $M$ spin-$J$ sensors jointly estimating the source strength and position near $(\beta_0,\xi_0)$. (b) Joint spectrum for five spin-1 nodes at $\alpha=2$ and $W=I_2$, shown with $\vec{\boldsymbol{v}}_M$ horizontal. The dashed blue minimum enclosing ball (MEB) contacts only the all-aligned pair $\pm\vec{\boldsymbol{v}}_M$. The solid red optimal ellipse $\mathcal E^\star$ also contacts $\pm\vec{\boldsymbol{v}}_1$, where the nearest node is aligned opposite to the others. The enlarged view resolves these four contacts. Light-blue dots are nonvertex spectral points, open gray circles are noncontact vertices, and filled circles are ellipse contacts. Blue-filled circles are also MEB contacts.}
	\label{fig:source_application}
\end{figure}

Motivated by magnetic-source localization from spatial field measurements~\cite{WengEtAl2026NatCommun,MarasliEtAl2026PRApplied}, we consider the joint estimation of the strength and position of a localized source. Consider a field source located at $\xi$ with the strength $\beta$, where \((\xi,\beta)\) is in the neighborhood of known values $(\xi_0,\beta_0)$. As shown in Fig.~\ref{fig:source_application}(a), $M\ge2$ spin-$J$ sensors, located at $x_k=\xi_0+kd$ with $k=1, \cdots, M$, are employed to estimate \((\xi,\beta)\). The dynamics of the $M$ sensors interacting with the field generated by the source can be described by the Hamiltonian
\begin{equation}
	\mathcal H_{\mathrm{src}}(\beta,\xi)
	=\sum_{k=1}^M\frac{\beta}{(x_k-\xi)^\alpha}J_z^{(k)},
	\label{eq:source_hamiltonian}
\end{equation}
where $\alpha\ge2$ is the degree of decay for the field strength with respect to the distance and is assumed to be known. For example, $\alpha=2$ for the electric field, which decays quadratically with the distance to the source. 

For an interrogation time $t$, we estimate deviations from the calibrated operating point $(\beta_0,\xi_0)$, with $\beta_0\ne0$. We use the dimensionless parameters $\varphi_\beta=t(\beta-\beta_0)/d^\alpha$ and $\varphi_\xi=t\beta_0(\xi-\xi_0)/d^{\alpha+1}$, and take $W=I_2$ to weight their estimation errors equally. After canceling the known reference Hamiltonian, the local generators are
\begin{equation}
	G_{\varphi_\beta}=\sum_{k=1}^M k^{-\alpha}J_z^{(k)},
	\qquad
	G_{\varphi_\xi}=\alpha\sum_{k=1}^M k^{-\alpha-1}J_z^{(k)}.
	\label{eq:source_generators}
\end{equation}
Appendix~\ref{app:source_encoding} derives these generators and relates the dimensionless cost to the physical estimation errors.

The common eigenstates of the two generators are products of local $J_z$ eigenstates, $|s_1s_2\ldots s_M\rangle$ with $s_k\in\{-J,-J+1,\ldots,J\}$, $k=1,\cdots, M$. The associated eigenvalue vectors are
\begin{equation}
	\vec{\boldsymbol{g}}_i=\sum_{k=1}^M s_k k^{-\alpha}\left(1,\frac{\alpha}{k}\right)^{\mathsf T}.
	\label{eq:source_eigenvalue_vector}
\end{equation}
These eigenvalue vectors are distributed symmetrically about the origin, i.e., if $\vec{\boldsymbol{g}}_i$ is an eigenvalue vector associated with the eigenstate $|s_1s_2\ldots s_M\rangle$, then $-\vec{\boldsymbol{g}}_i$ is also an eigenvalue vector, associated with the eigenstate $|(-s_1)(-s_2)\ldots (-s_M)\rangle$. For such a set of eigenvalue vectors, the minimal enclosing ball or ellipsoid is always centered at the origin.

The farthest pair from the origin corresponds to the all-aligned states $|\pm J\rangle^{\otimes M}$, with
\begin{equation}
	\vec{\boldsymbol{v}}_M=J\left(\sum_{k=1}^M k^{-\alpha},\alpha\sum_{k=1}^M k^{-\alpha-1}\right)^{\mathsf T}.
	\label{eq:source_aligned_vertex}
\end{equation}
The minimum enclosing ball therefore contacts only $\pm\vec{\boldsymbol{v}}_M$, as shown in Fig.~\ref{fig:source_application}(b), and gives
\begin{equation}
	\max_{|\psi_0\rangle}\operatorname{Tr}[F_Q(|\psi_0\rangle)]=4|\vec{\boldsymbol{v}}_M|^2.
	\label{eq:source_maximum_total_qfi}
\end{equation}
This value is attained by $|C_M\rangle=(|+J\rangle^{\otimes M}+e^{i\chi_M}|-J\rangle^{\otimes M})/\sqrt2$. The two occupied points are shown as the blue contacts on a diameter in Fig.~\ref{fig:source_application}(b).

To minimize the weighted covariance, we determine the optimal enclosing ellipse. Besides $\pm\vec{\boldsymbol{v}}_M$, consider the pair $\pm\vec{\boldsymbol{v}}_1$, where $\vec{\boldsymbol{v}}_1=2J(1,\alpha)^{\mathsf T}-\vec{\boldsymbol{v}}_M$ is the eigenvalue vector associated with the eigenstate $|+J\rangle_1\otimes|-J\rangle^{\otimes(M-1)}$, i.e., the first node in $|+J\rangle$ state and the rest in $|-J\rangle$ state. We first find the ellipse enclosing these four points with the smallest harmonic mean of its semiaxes. Let $B=(\vec{\boldsymbol{v}}_1\ \vec{\boldsymbol{v}}_M)$ be a $2\times 2$ matrix and $\eta=\vec{\boldsymbol{v}}_1^{\mathsf T}\vec{\boldsymbol{v}}_M/(|\vec{\boldsymbol{v}}_1|\,|\vec{\boldsymbol{v}}_M|)$, the optimal ellipse is given by $\mathcal E^\star=\{\vec{\boldsymbol{g}}:\vec{\boldsymbol{g}}^{\mathsf T}S^\star\vec{\boldsymbol{g}}\le R_{S^\star}^2 \}$ with $R_{S^\star}=1$ and
\begin{equation}
	S^\star=(B^{-1})^{\mathsf T}\begin{pmatrix}1&-\eta\\-\eta&1\end{pmatrix}B^{-1}.
	\label{eq:source_ellipse_matrix}
\end{equation}
For the array considered here, all remaining eigenvalue vectors lie strictly inside this ellipse. It is therefore also optimal for the full joint spectrum, with exactly the four contacts, as shown in the enlarged view of Fig.~\ref{fig:source_application}(b). The detailed proofs can be found in Appendix~\ref{app:source_ellipse}. The optimal QFIM, which can be obtained by Eq.~\eqref{eq:targetqfim}, yields
\begin{equation}
	F_Q^\star=\frac{4}{|\vec{\boldsymbol{v}}_1|+|\vec{\boldsymbol{v}}_M|}\left(|\vec{\boldsymbol{v}}_M|\vec{\boldsymbol{v}}_1\vec{\boldsymbol{v}}_1^{\mathsf T}+|\vec{\boldsymbol{v}}_1|\vec{\boldsymbol{v}}_M\vec{\boldsymbol{v}}_M^{\mathsf T}\right).
	\label{eq:source_target_qfim}
\end{equation}

We now identify the optimal probe state, which has support on the eigenstates associated with the four contact points marked by filled circles in Fig.~\ref{fig:source_application}(b). Since $\vec{\boldsymbol{v}}_1$ and $\vec{\boldsymbol{v}}_M$ are linearly independent, the zero-centroid condition requires equal populations within each opposite pair. Matching Eq.~\eqref{eq:source_target_qfim} then gives the total populations $p_1^\star=|\vec{\boldsymbol{v}}_M|/(|\vec{\boldsymbol{v}}_1|+|\vec{\boldsymbol{v}}_M|)$ and $p_M^\star=|\vec{\boldsymbol{v}}_1|/(|\vec{\boldsymbol{v}}_1|+|\vec{\boldsymbol{v}}_M|)$. Denote $|D_1\rangle=|+J\rangle_1\otimes|-J\rangle^{\otimes(M-1)}$, $|\overline D_1\rangle=|-J\rangle_1\otimes|+J\rangle^{\otimes(M-1)}$, $|C_1\rangle=(|D_1\rangle+e^{i\chi_1}|\overline D_1\rangle)/\sqrt2$ and $|C_M\rangle=(|+J\rangle^{\otimes M}+e^{i\chi_M}|-J\rangle^{\otimes M})/\sqrt2$, the optimal probes then take the form
\begin{equation}
	|\psi_0^\star\rangle=\sqrt{\frac{|\vec{\boldsymbol{v}}_M|}{|\vec{\boldsymbol{v}}_1|+|\vec{\boldsymbol{v}}_M|}}\,|C_1\rangle+e^{i\chi}\sqrt{\frac{|\vec{\boldsymbol{v}}_1|}{|\vec{\boldsymbol{v}}_1|+|\vec{\boldsymbol{v}}_M|}}\,|C_M\rangle,
	\label{eq:source_optimum}
\end{equation}
where the phases can be arbitrarily chosen.
The corresponding minimal weighted covariance is
\begin{equation}
	C_{I_2}^\star=\frac{(|\vec{\boldsymbol{v}}_1|+|\vec{\boldsymbol{v}}_M|)^2}{4|\det(\vec{\boldsymbol{v}}_1,\vec{\boldsymbol{v}}_M)|^2}.
	\label{eq:source_precision_lower_bound}
\end{equation}
The detailed derivation can be found in Appendix~\ref{app:source_ellipse}, where a projective measurement that attains $F_Q^\star$ at the operating point is also provided.

For single-parameter estimation, the optimal probe state is $|C_M\rangle$ for both parameters. However, the QFIM of this state is $4\vec{\boldsymbol{v}}_M\vec{\boldsymbol{v}}_M^{\mathsf T}$, which is singular, thus can not resolve both parameters. The joint estimation thus requires the additional pair $\pm\vec{\boldsymbol{v}}_1$. In fact, since $|\vec{\boldsymbol{v}}_M|>|\vec{\boldsymbol{v}}_1|$, the optimal state in Eq.(\ref{eq:source_optimum}) assigns a larger population to $|C_1\rangle$, which is not optimal for estimating either parameter separately. Under the stated array conditions, the geometry selects only four common eigenstates for any $M\ge2$, and fixes their populations analytically.

To quantify the gain from entanglement between nodes, we compare the optimal cost with the minimum attainable using node-separable probes. As shown in Appendix~\ref{app:source_benchmark}, the optimal product state is $\bigotimes_{k=1}^M\bigl[(|+J\rangle_k+e^{i\chi_k}|-J\rangle_k)/\sqrt2\bigr]$, which gives
\begin{equation}
	C_{I_2,\mathrm{sep}}=\frac1{4J^2}\operatorname{Tr}\left[\left(\sum_{k=1}^M\vec{\boldsymbol r}_k\vec{\boldsymbol r}_k^{\mathsf T}\right)^{-1}\right],
	\label{eq:source_separable_cost}
\end{equation}
where $\vec{\boldsymbol r}_k=k^{-\alpha}(1,\alpha/k)^{\mathsf T}$.
For five spin-1 nodes with $\alpha=2$ in Fig.~\ref{fig:source_application}, $C_{I_2}^\star\simeq4.10$ and $C_{I_2,\mathrm{sep}}\simeq13.20$. The optimal probe therefore reduces the attainable weighted covariance to about $31\%$ of the node-separable minimum.

As the array grows at fixed node spacing $d$ and spin $J$, the contributions from increasingly distant sensors decrease. Writing the node number explicitly, for $\alpha\ge2$ we obtain
\begin{equation}
	C_{I_2,M}^\star=C_{I_2,\infty}^\star+O(M^{1-\alpha}).
	\label{eq:source_fixed_spacing_limit}
\end{equation}
The cost therefore approaches a positive constant, so adding more distant nodes eventually gives little further improvement. Appendix~\ref{app:source_scaling} derives this limit.

\section{Summary}\label{sec:summary}

We have extended the spectral construction of optimal probes from single-parameter estimation to joint estimation with commuting generators. The two spectral endpoints are replaced by contact points of an optimal ellipsoid enclosing the joint spectrum. These points identify the common eigenstates that can contribute to an optimal probe, while the ellipsoid center and shape specify the centroid and QFIM that their populations must reproduce. The resulting conditions characterize all optimal pure probes and determine the minimum attainable weighted covariance.

The geometry provides analytic constructions for several spectral configurations. For two parameters, the minimum weighted covariance is determined by the smallest harmonic mean of the enclosing ellipse's semiaxes in the coordinates set by the weight matrix. When the spectral convex hull is a triangle, the optimal ellipse is centered at its incenter, and the optimal populations are proportional to the opposite side lengths. The precision is then fixed by the triangle's inradius. When an analytic construction is unavailable, the semidefinite formulation provides a complementary numerical method for finding the ellipse, followed by the same population conditions.

The distributed sensing applications show how the estimation task determines which states must be populated. For a field mean and gradient, changing the covariance weight redistributes populations between the two pairs of common eigenstates. The localized-source example reveals a different obstruction: the same all-aligned superposition is optimal for estimating either source strength or position separately, but it encodes only one local combination of the two parameters. Joint estimation requires additional configurations in which the nearest node is aligned opposite to the others. For the array considered here, the optimal probes retain support on only four common eigenstates as the number of nodes grows. Their populations and an attaining measurement are obtained analytically, providing an explicit sensing protocol and a quantitative comparison with node-separable probes.

Several directions remain open. Extending the construction beyond commuting generators, where the joint spectrum is no longer a fixed finite set and the ellipsoid picture must be replaced by a more general operator-geometric framework, is a natural next step. Another direction is to extend the construction to noisy evolution, where the dynamics is no longer unitary. A promising route is the purification approach, in which the system is embedded in a larger purified space and the noise is represented as unitary evolution on the enlarged Hilbert space. 

\begin{acknowledgments}
This work is supported by the National Natural Science Foundation of China (Grants No. 12505024, No. 92476201), Department of Science and Technology of Guangdong Province (Grant No. 2024QN11X234), Guangdong Basic and Applied Basic Research Foundation (Grant No. 2025A1515011441), Shenzhen Science and Technology Program (Grants No. ZDYJ20251211120900001, No. JCYJ20240813141350066),  Quantum Science and Technology-National Science and Technology Major Project (Grant No. 2023ZD0300600), the Research Grants Council of Hong Kong (Grants No. 14309223, No. 14309624, No. 14309022), Guangdong Provincial Quantum Science Strategic Initiative (Grants No. GDZX2303007, No. GDZX2505003), 1+1+1 CUHK-CUHK(SZ)-GDST Joint Collaboration Fund (Grant No. GRDP2025-022).
\end{acknowledgments}

\appendix

\section{Discussions on singular QFIMs and weight matrices}\label{app:rank}

For a unit vector $\bm u$, consider the parameter combination $\varphi_{\bm u}=\sum_{j=1}^m u_j\varphi_j$. With all orthogonal parameter combinations held fixed, the QFI for estimating $\varphi_{\bm u}$ is $\bm u^T F_Q(|\psi_p\rangle)\bm u$. For commuting generators,
\begin{equation}
	\frac14\bm u^T F_Q\left(\left|\psi_p\right\rangle\right)\bm u=\sum_{i:p_i>0}p_i\left[\bm u^T\left(\vec{\boldsymbol{g}}_i-\overline{\bm g}_p\right)\right]^2.
	\label{eq:rankvariance}
\end{equation}
Since every occupied point has $p_i>0$, the right-hand side of Eq.~\eqref{eq:rankvariance} vanishes if and only if $\bm u^T(\vec{\boldsymbol{g}}_i-\overline{\bm g}_p)=0$ for every occupied point. These conditions imply $\bm u^T(\vec{\boldsymbol{g}}_i-\vec{\boldsymbol{g}}_j)=0$ for every occupied pair. Conversely, if all these pairwise differences are orthogonal to $\bm u$, their projections are identical and equal to the projection of their centroid.

Thus the directions with zero QFI are precisely those perpendicular to every difference between occupied spectral points. Since the QFIM is positive semidefinite, these directions form its kernel:
\begin{equation}
	\ker F_Q(|\psi_p\rangle)=\left(\operatorname{span}\left\{\vec{\boldsymbol{g}}_i-\vec{\boldsymbol{g}}_j:p_ip_j>0\right\}\right)^\perp.
\end{equation}
Taking dimensions gives
\begin{equation}
	\begin{aligned}
		\operatorname{rank}F_Q(|\psi_p\rangle)
		&=m-\dim\ker F_Q(|\psi_p\rangle)\\
		&=\dim\operatorname{span}\left\{\vec{\boldsymbol{g}}_i-\vec{\boldsymbol{g}}_j:p_ip_j>0\right\}.
	\end{aligned}
\end{equation}
A probe with a singular QFIM cannot support locally unbiased estimation of all parameters. We therefore assign it infinite cost when $W\succ0$, which gives nonzero weight to every parameter direction.

The conclusion of infinite cost above assumes $W\succ0$. If some weights vanish, the corresponding estimation errors no longer enter the cost. The associated parameters, however, can still affect the precision of the parameters of interest if their values are unknown~\cite{DemkowiczDobrzanskiGoreckiGuta2020,SuzukiYangHayashi2020,TsangAlbarelliDatta2020}. To see this, choose coordinates with $W=\operatorname{diag}(W_a,0)$, where $W_a\succ0$, and denote the parameters of interest by $\bm{\varphi}_a$ and the remaining unknown parameters by $\bm{\varphi}_n$. For $F_Q\succ0$, write its corresponding blocks as $F_{aa},F_{an},F_{na},F_{nn}$. Block inversion gives
\begin{equation}
	\operatorname{Tr}(WF_Q^{-1})=\operatorname{Tr}\left[W_a\left(F_{aa}-F_{an}F_{nn}^{-1}F_{na}\right)^{-1}\right].
	\label{eq:nuisance}
\end{equation}
The Schur complement accounts for the uncertainty in the nuisance parameters $\bm{\varphi}_n$. If these parameters are known, the estimation problem uses $F_{aa}$ alone. If they are unknown, the same simplification holds only when $F_{an}=0$.

\section{Achievability of the precision bound}\label{app:matrix}

In this appendix, we prove that the lower bound in Eq.~\eqref{eq:geometriclowerbound}, given by
\begin{equation}
	C_W^\star\geq\max_{S\succeq0,\,S\ne0}\frac{\left(\operatorname{Tr}\sqrt{S^{\frac12}WS^{\frac12}}\right)^2}{4R_S^2},
	\label{eq:appendixlowerbound}
\end{equation}
 is tight. We then derive the conditions that an optimal probe must satisfy, which are used in the main text to construct explicit optimal probe states. We also give a short proof of the matrix Cauchy--Schwarz inequality that underlies this bound.

\subsection{Saturation of the precision bound}\label{app:populationvariation}

The lower bound obtained from Eq.~\eqref{eq:Sbound} is attained when the same probe saturates both inequalities. We show that a probe minimizing the weighted cost does so for the choice $S^\star=(F_Q^\star)^{-1}W(F_Q^\star)^{-1}$ in Eq.~\eqref{eq:sstar}. This fixes the scale only for the proof. Positive rescaling leaves the enclosing ellipsoid and the target QFIM in Eq.~\eqref{eq:targetqfim} unchanged.

Under our assumption on the joint spectrum, choosing $p_i>0$ for every spectral point gives a nonsingular QFIM and hence a finite cost. By contrast, the cost diverges whenever an eigenvalue of the QFIM approaches zero, since
\begin{equation}
	\operatorname{Tr}\left[WF_Q(|\psi_p\rangle)^{-1}\right]\geq\frac{\lambda_{\min}(W)}{\lambda_{\min}[F_Q(|\psi_p\rangle)]}.
	\label{eq:smcoercive}
\end{equation}
Since the normalized populations form a compact set, and the cost is continuous wherever the QFIM is nonsingular, the minimum is therefore attained at a nonsingular QFIM.

Let $p^\star$ be an optimal population distribution, with $F_Q^\star=F_Q(|\psi_{p^\star}\rangle)\succ0$, and define $S^\star$ as in Eq.~\eqref{eq:sstar}. For any population distribution $q$, consider $p_t=(1-t)p^\star+tq$ with $0\leq t\leq1$. The centroid corresponding to $p_t$ is
\begin{equation}
	\overline{\bm g}_{p_t}=(1-t)\overline{\bm g}_{p^\star}+t\overline{\bm g}_q.
	\label{eq:smpopulationmean}
\end{equation}
Since $\sum_i p_i\vec{\boldsymbol{g}}_i\vec{\boldsymbol{g}}_i^T$ is linear in the populations, the covariance expression in Eq.~\eqref{eq:covariance} gives
\begin{equation}
	\begin{aligned}
		&\frac14\left[F_Q(|\psi_{p_t}\rangle)-(1-t)F_Q^\star-tF_Q(|\psi_q\rangle)\right]\\
		&=(1-t)\overline{\bm g}_{p^\star}\overline{\bm g}_{p^\star}^T+t\overline{\bm g}_q\overline{\bm g}_q^T-\overline{\bm g}_{p_t}\overline{\bm g}_{p_t}^T=t(1-t)\bm d\bm d^T,
	\end{aligned}
	\label{eq:smqfimvariation}
\end{equation}
where $\bm d=\overline{\bm g}_{p^\star}-\overline{\bm g}_q$. Equivalently,
\begin{equation}
	F_Q(|\psi_{p_t}\rangle)=(1-t)F_Q^\star+tF_Q(|\psi_q\rangle)+4t(1-t)\bm d\bm d^T.
	\label{eq:qfimconcavity}
\end{equation}
The additional term comes from the difference between the two centroids.

To differentiate the cost, write $F(t)=F_Q(|\psi_{p_t}\rangle)$. Equation~\eqref{eq:qfimconcavity} gives $F(t)\succeq(1-t)F_Q^\star\succ0$ for $0\leq t<1$, so the inverse exists even if $F_Q(|\psi_q\rangle)$ is singular. Differentiating $F(t)F(t)^{-1}=I$ yields
\begin{equation}
	\frac{dF(t)^{-1}}{dt}=-F(t)^{-1}\frac{dF(t)}{dt}F(t)^{-1}.
	\label{eq:sminversederivative}
\end{equation}
From Eq.~\eqref{eq:qfimconcavity},
\begin{equation}
	\left.\frac{dF(t)}{dt}\right|_{t=0^+}=F_Q(|\psi_q\rangle)-F_Q^\star+4\bm d\bm d^T.
	\label{eq:smqfimderivative}
\end{equation}
The cyclic property of the trace and the definition of $S^\star$ give
\begin{equation}
	\left.\frac{d}{dt}\operatorname{Tr}[WF(t)^{-1}]\right|_{t=0^+}=-\operatorname{Tr}\left[S^\star\left.\frac{dF(t)}{dt}\right|_{t=0^+}\right].
	\label{eq:smcostvariation}
\end{equation}
Since $p^\star$ minimizes the cost, the right derivative is nonnegative. Substituting Eq.~\eqref{eq:smqfimderivative} yields
\begin{equation}
	\begin{aligned}
		0&\leq\left.\frac{d}{dt}\operatorname{Tr}[WF(t)^{-1}]\right|_{t=0^+}\\
		&=\operatorname{Tr}(S^\star F_Q^\star)-\operatorname{Tr}[S^\star F_Q(|\psi_q\rangle)]-4\bm d^TS^\star\bm d.
	\end{aligned}
	\label{eq:weightedmaxproof}
\end{equation}
Rearranging gives
\begin{equation}
	\begin{aligned}
		\operatorname{Tr}[S^\star F_Q(|\psi_q\rangle)]
		&\leq\operatorname{Tr}(S^\star F_Q^\star)-4\bm d^TS^\star\bm d\\
		&\leq\operatorname{Tr}(S^\star F_Q^\star).
	\end{aligned}
	\label{eq:smweightedqfi}
\end{equation}
Because Eq.~\eqref{eq:smweightedqfi} holds for every population distribution $q$, $p^\star$ maximizes the weighted total QFI for $S^\star$. Equation~\eqref{eq:weightedmeb} therefore gives $4R_{S^\star}^2=\operatorname{Tr}(S^\star F_Q^\star)$. Substituting $S^\star=(F_Q^\star)^{-1}W(F_Q^\star)^{-1}$, we obtain
\begin{equation}
	\operatorname{Tr}(S^\star F_Q^\star)=\operatorname{Tr}\left[W(F_Q^\star)^{-1}\right]=C_W^\star.
\end{equation}

The numerator can be evaluated directly from the same choice of $S^\star$. The matrices $(S^\star)^{1/2}W(S^\star)^{1/2}$ and $W^{1/2}S^\star W^{1/2}$ have the same eigenvalues, and
\[
	W^{1/2}S^\star W^{1/2}=\left[W^{1/2}(F_Q^\star)^{-1}W^{1/2}\right]^2.
\]
Since $W^{1/2}(F_Q^\star)^{-1}W^{1/2}$ is positive definite, taking the positive square root gives
\begin{equation}
	\begin{aligned}
		\operatorname{Tr}\sqrt{S^{\star\frac12}WS^{\star\frac12}}&=\operatorname{Tr}\left[W^{1/2}(F_Q^\star)^{-1}W^{1/2}\right]\\
		&=\operatorname{Tr}\left[W(F_Q^\star)^{-1}\right]=C_W^\star.
	\end{aligned}
\end{equation}
Combining these results, we have
\begin{equation}
	4R_{S^\star}^2=\operatorname{Tr}(S^\star F_Q^\star)=\operatorname{Tr}\sqrt{S^{\star\frac12}WS^{\star\frac12}}=C_W^\star,
	\label{eq:smprecisionsaturation}
\end{equation}
which further gives $\left(\operatorname{Tr}\sqrt{S^{\star\frac12}WS^{\star\frac12}}\right)^2/(4R_{S^\star}^2)=C_W^\star$ and establishes Eq.~\eqref{eq:geometriclowerbound}.

The equality conditions also characterize all optimal pure probes. For any optimal population distribution $p$, write $F=F_Q(|\psi_p\rangle)$ and keep the same $S^\star$. Equations~\eqref{eq:Sbound} and \eqref{eq:smprecisionsaturation} give
\begin{equation}
	\begin{aligned}
		C_W^\star&=\operatorname{Tr}(WF^{-1})\\
		&\geq\frac{\left(\operatorname{Tr}\sqrt{S^{\star\frac12}WS^{\star\frac12}}\right)^2}{\operatorname{Tr}(S^\star F)}\\
		&\geq\frac{\left(\operatorname{Tr}\sqrt{S^{\star\frac12}WS^{\star\frac12}}\right)^2}{4R_{S^\star}^2}=C_W^\star.
	\end{aligned}
	\label{eq:smmomentuniquechain}
\end{equation}
The geometric equality restricts the occupied points to $I^\star$ and places their centroid at $\vec{\boldsymbol{c}}^\star$. The matrix equality gives $S^\star=\gamma F^{-1}WF^{-1}$, while $\operatorname{Tr}(S^\star F)=C_W^\star=\operatorname{Tr}(WF^{-1})$ fixes $\gamma=1$ for the scale chosen in this proof. Thus $FS^\star F=W$, whose unique positive-definite solution is the target QFIM in Eq.~\eqref{eq:targetqfim}, with $4R_{S^\star}^2/\bigl(\operatorname{Tr}\sqrt{S^{\star\frac12}WS^{\star\frac12}}\bigr)=1$. Every optimal population therefore satisfies Eq.~\eqref{eq:momentmatching}. Conversely, any nonnegative solution has QFIM $F_Q^\star$ and cost $C_W^\star$. Positive rescaling of $S^\star$ leaves these conditions unchanged. All optimal populations share the same centroid and QFIM, while the populations and relative phases need not be unique. Within a degenerate joint eigenspace, only the total population is constrained.

\subsection{Matrix Cauchy--Schwarz inequality}\label{app:matrixcs}

For $F,W\succ0$ and $S\succeq0$, let $X=(W^{1/2}SW^{1/2})^{1/2}$ and $K=W^{-1/2}FW^{-1/2}$. Since $S^{1/2}WS^{1/2}$ and $W^{1/2}SW^{1/2}$ have the same eigenvalues, $\operatorname{Tr}\sqrt{S^{\frac12}WS^{\frac12}}=\operatorname{Tr}X$. Applying Cauchy--Schwarz to the matrices $K^{-1/2}$ and $K^{1/2}X$ gives
\begin{equation}
	\begin{aligned}
		\left(\operatorname{Tr}\sqrt{S^{\frac12}WS^{\frac12}}\right)^2&=(\operatorname{Tr}X)^2\\
		&\leq\operatorname{Tr}(K^{-1})\operatorname{Tr}(XKX)\\
		&=\operatorname{Tr}(WF^{-1})\operatorname{Tr}(SF),
	\end{aligned}
	\label{eq:smcsproof}
\end{equation}
which proves Eq.~\eqref{eq:matrixcs}. For nonzero $S$, equality holds precisely when $K^{1/2}X=\alpha K^{-1/2}$ for some $\alpha>0$. Substituting $X=\alpha K^{-1}$ into $S=W^{-1/2}X^2W^{-1/2}$ gives $S=\alpha^2F^{-1}WF^{-1}$, establishing the equality condition used in the precision bound.

\section{Semidefinite formulations}\label{app:sdp}

The SDPs in this appendix complement the geometric construction. We derive the enclosing-ellipsoid program in Eq.~\eqref{eq:ellipsoid_sdp} from the precision bound and explain how its solution determines an optimal probe. We then give an equivalent program that optimizes the populations directly, together with upper and lower bounds for checking the reconstructed precision.

\subsection{The enclosing-ellipsoid SDP}\label{app:enclosingsdp}

The ratio in Eq.~\eqref{eq:geometriclowerbound} is unchanged under $S\mapsto\gamma S$, since $\operatorname{Tr}\sqrt{(\gamma S)^{\frac12}W(\gamma S)^{\frac12}}=\sqrt\gamma\,\operatorname{Tr}\sqrt{S^{\frac12}WS^{\frac12}}$ and $R_{\gamma S}^2=\gamma R_S^2$. We can therefore fix the scale by imposing $R_S^2\leq1$ and maximize $\operatorname{Tr}\sqrt{S^{\frac12}WS^{\frac12}}$ instead. At an optimum, $R_S^2=1$: otherwise $S$ could be increased by a positive factor to raise the objective while preserving the constraint. Thus
\begin{equation}
	C_W^\star=\frac14\left[\max_{S\succeq0,\,R_S^2\leq1}\operatorname{Tr}\sqrt{S^{\frac12}WS^{\frac12}}\right]^2.
	\label{eq:normalizedprecision}
\end{equation}

To express the objective as an SDP, we use
\begin{equation}
	\operatorname{Tr}\sqrt{S^{\frac12}WS^{\frac12}}=\max_{X\in\mathbb R^{m\times m}}\left\{\operatorname{Tr}X:\begin{pmatrix}S&X\\X^T&W\end{pmatrix}\succeq0\right\}.
	\label{eq:fidelitysdp}
\end{equation}
Here $X$ need not be symmetric. The block constraint is equivalent to $X=S^{1/2}VW^{1/2}$ for a real matrix $V$ with $\|V\|_{\mathrm{op}}\leq1$. Maximizing its trace gives the trace norm of $W^{1/2}S^{1/2}$, which equals $\operatorname{Tr}\sqrt{S^{\frac12}WS^{\frac12}}$. This representation also holds when $S$ is singular.

The remaining constraint, $R_S^2\leq1$, requires a center $\vec{\boldsymbol{c}}$ such that $(\vec{\boldsymbol{g}}_i-\vec{\boldsymbol{c}})^TS(\vec{\boldsymbol{g}}_i-\vec{\boldsymbol{c}})\leq1$ for every spectral point. Expanding this expression introduces the products $S\vec{\boldsymbol{c}}$ and $\vec{\boldsymbol{c}}^TS\vec{\boldsymbol{c}}$, which are nonlinear in the optimization variables. We instead introduce $\bm q=S\vec{\boldsymbol{c}}$ and $t=\vec{\boldsymbol{c}}^TS\vec{\boldsymbol{c}}$, so that the enclosing inequalities become
\begin{equation}
	\vec{\boldsymbol{g}}_i^TS\vec{\boldsymbol{g}}_i-2\bm q^T\vec{\boldsymbol{g}}_i+t\leq1\qquad\text{for all }i.
	\label{eq:enclosingaffine}
\end{equation}
The generalized Schur complement imposes consistency with an enclosing center:
\begin{equation}
	\begin{gathered}
		\begin{pmatrix}S&\bm q\\\bm q^T&t\end{pmatrix}\succeq0\\
		\Longleftrightarrow\quad S\succeq0,\quad\bm q\in\operatorname{range}(S),\quad t\geq\bm q^TS^+\bm q.
	\end{gathered}
	\label{eq:smgeneralizedschur}
\end{equation}
Indeed, choosing $\vec{\boldsymbol{c}}=S^+\bm q$ gives
\begin{equation}
	\begin{aligned}
		(\vec{\boldsymbol{g}}_i-\vec{\boldsymbol{c}})^TS(\vec{\boldsymbol{g}}_i-\vec{\boldsymbol{c}})
		&=\vec{\boldsymbol{g}}_i^TS\vec{\boldsymbol{g}}_i-2\bm q^T\vec{\boldsymbol{g}}_i+\bm q^TS^+\bm q\\
		&\leq\vec{\boldsymbol{g}}_i^TS\vec{\boldsymbol{g}}_i-2\bm q^T\vec{\boldsymbol{g}}_i+t\leq1.
	\end{aligned}
	\label{eq:smcompletequadratic}
\end{equation}
Conversely, any center satisfying the enclosing condition gives feasible variables $\bm q=S\vec{\boldsymbol{c}}$ and $t=\vec{\boldsymbol{c}}^TS\vec{\boldsymbol{c}}$. Thus Eqs.~\eqref{eq:enclosingaffine} and \eqref{eq:smgeneralizedschur} are an exact representation of $R_S^2\leq1$, including for singular $S$.

Combining these enclosing constraints with Eq.~\eqref{eq:fidelitysdp} gives the SDP in Eq.~\eqref{eq:ellipsoid_sdp}, with objective $\operatorname{Tr}X$ and cost determined by Eq.~\eqref{eq:normalizedprecision}.

\subsection{Reconstructing optimal probes}\label{app:sdpreconstruction}

Let $S^\star,X^\star,\bm q^\star,t^\star$ solve Eq.~\eqref{eq:ellipsoid_sdp}. At the optimum, $R_{S^\star}^2=1$ and $\operatorname{Tr}X^\star=\operatorname{Tr}\sqrt{S^{\star\frac12}WS^{\star\frac12}}$, so
\begin{equation}
	C_W^\star=\frac{[\operatorname{Tr}X^\star]^2}{4}=\frac{\left(\operatorname{Tr}\sqrt{S^{\star\frac12}WS^{\star\frac12}}\right)^2}{4}.
	\label{eq:sdpoptimalvalues}
\end{equation}
The equality condition of the precision bound gives $S^\star=\gamma(F_Q^\star)^{-1}W(F_Q^\star)^{-1}$ for some $\gamma>0$, so $S^\star\succ0$. The ellipsoid center is therefore $\vec{\boldsymbol{c}}^\star=(S^\star)^{-1}\bm q^\star$, and its contact points satisfy $(\vec{\boldsymbol{g}}_i-\vec{\boldsymbol{c}}^\star)^TS^\star(\vec{\boldsymbol{g}}_i-\vec{\boldsymbol{c}}^\star)=1$.

The normalization $R_{S^\star}^2=1$ generally differs from the choice $\gamma=1$ in Eq.~\eqref{eq:sstar}. To obtain the QFIM from the SDP output, solve the matrix equality as $F_Q^\star=\sqrt\gamma\,W^{1/2}(W^{1/2}S^\star W^{1/2})^{-1/2}W^{1/2}$. Its weighted cost is $\operatorname{Tr}\sqrt{S^{\star\frac12}WS^{\star\frac12}}/\sqrt\gamma$. Comparing with the geometric precision bound fixes $\sqrt\gamma=4R_{S^\star}^2/\bigl(\operatorname{Tr}\sqrt{S^{\star\frac12}WS^{\star\frac12}}\bigr)$ and gives
\begin{equation}
	F_Q^\star=\frac{4R_{S^\star}^2}{\operatorname{Tr}\sqrt{S^{\star\frac12}WS^{\star\frac12}}}W^{1/2}\left(W^{1/2}S^\star W^{1/2}\right)^{-1/2}W^{1/2}.
	\label{eq:scaletarget}
\end{equation}
For the SDP output, $R_{S^\star}^2=1$, but Eq.~\eqref{eq:scaletarget} also holds after positive rescaling, as in Eq.~\eqref{eq:targetqfim}. With its center, contact points, and target QFIM determined, the nonnegative populations follow from Eq.~\eqref{eq:momentmatching}. Every solution gives an optimal pure probe, with arbitrary relative phases.

\subsection{Direct population optimization and precision checks}\label{app:populationsdp}

We provide an alternative SDP determining the populations, which provides an independent check of the geometric reconstruction. 

For the probe state, $|\psi_p\rangle = \sum_{k=1}^d \sqrt{p_k} e^{i\phi_k}|k\rangle$, the QFIM is given by 
\begin{equation}
F_Q = 4\sum_{k=1}^d p_k (\vec{\boldsymbol{g}}_k - \bar{\boldsymbol{g}})(\vec{\boldsymbol{g}}_k - \bar{\boldsymbol{g}})^{\mathsf{T}}.
\end{equation} To express the QFIM using linear matrix constraints, define $\bm z_k=(\vec{\boldsymbol{g}}_k^T,1)^T$ and
\begin{equation}
	 M(p)=\sum_k p_k\bm z_k\bm z_k^T=
\begin{pmatrix}
\sum_k p_k \vec{\boldsymbol{g}}_k \vec{\boldsymbol{g}}_k^{\mathsf{T}} & \bar{\boldsymbol{g}} \\
\bar{\boldsymbol{g}}^{\mathsf{T}} & 1
\end{pmatrix}.
	\label{eq:augmentedmoments}
\end{equation}
The inverse of $M(p)$ can be written as
\begin{equation}
	 M(p)^{-1}=
\begin{pmatrix}
C(p)^{-1} & * \\
* & *
\end{pmatrix},
\end{equation}
where $C(p)=\sum_k p_k \vec{\boldsymbol{g}}_k \vec{\boldsymbol{g}}_k^{\mathsf{T}}-\bar{\boldsymbol{g}}\bar{\boldsymbol{g}}^{\mathsf{T}}$ is the Schur complement of the bottom block, which equals $F_Q(|\psi_p\rangle)/4 $. 
Let $E=(I_m,0_{m\times1})^T$, then 
\begin{equation}
E^{\mathsf{T}}M(p)^{-1} E = C(p)^{-1}.
\end{equation} 
We now introduce an auxiliary symmetric matrix $X$ and impose the linear matrix inequality
\begin{equation}
\begin{pmatrix}
M(p) & E \\
E^{\mathsf{T}} & X
\end{pmatrix}
\succeq 0.
\label{eq:main-LMI}
\end{equation}
By the generalized Schur complement lemma, this is equivalent to
\begin{equation}
X \succeq E^{\mathsf{T}} M(p)^{-1} E =  C(p)^{-1}.
\end{equation}
Since $F_Q = 4 C(p)$, we have $ C(p)^{-1} = 4 F_Q^{-1}$, so $X \succeq 4 F_Q^{-1}$. Consequently,
\begin{equation}
\frac{1}{4}\Tr(W X) \geq \Tr(W F_Q^{-1}) \geq C_W^\star,
\end{equation}
with equality at the optimum. The weighted-covariance optimization is therefore exactly equivalent to the SDP
\begin{equation}
\begin{aligned}
C_W^\star ={}&
\min_{p, X = X^{\mathsf{T}}}
\frac{1}{4}\Tr(W X) \\
\text{subject to }{}&
\sum_{k=1}^d p_k = 1,
\qquad p_k \geq 0, \\
&
\begin{pmatrix}
M(p) & E \\
E^{\mathsf{T}} & X
\end{pmatrix}
\succeq 0.
\end{aligned}
\label{eq:primal_sdp}
\end{equation}

The populations obtained by either route can be checked directly. For a population $p$ with a nonsingular QFIM, set $S_p=F_Q(|\psi_p\rangle)^{-1}WF_Q(|\psi_p\rangle)^{-1}$. This probe is optimal if and only if every spectral point satisfies
\begin{equation}
	4(\vec{\boldsymbol{g}}_i-\overline{\bm g}_p)^TS_p(\vec{\boldsymbol{g}}_i-\overline{\bm g}_p)\leq\operatorname{Tr}[WF_Q(|\psi_p\rangle)^{-1}].
	\label{eq:equivalencetest}
\end{equation}
The derivative of the cost toward a population concentrated at point $i$ is the right-hand side minus the left-hand side, proving necessity. For sufficiency, the population-weighted average of the left-hand side already equals the cost on the right. The inequalities therefore give an enclosing ellipsoid whose bound is attained by this probe, since the matrix Cauchy--Schwarz equality holds by the definition of $S_p$. Every occupied point must be a contact point.

For an approximate reconstruction, feasible populations and an enclosing ellipsoid give complementary bounds even when the equality conditions are not met. Let $\widehat p$ be normalized and nonnegative, with $\widehat F=F_Q(|\psi_{\widehat p}\rangle)\succ0$, and let nonzero $\widehat S\succeq0$, $\widehat{\vec{\boldsymbol{c}}}$, and $\widehat R>0$ satisfy $(\vec{\boldsymbol{g}}_i-\widehat{\vec{\boldsymbol{c}}})^T\widehat S(\vec{\boldsymbol{g}}_i-\widehat{\vec{\boldsymbol{c}}})\leq\widehat R^2$ for every spectral point. Then
\begin{equation}
	\operatorname{Tr}(W\widehat F^{-1})\geq C_W^\star\geq\frac{\left(\operatorname{Tr}\sqrt{\widehat S^{\frac12}W\widehat S^{\frac12}}\right)^2}{4\widehat R^2}.
	\label{eq:numericalcertificate}
\end{equation}
The difference between a feasible upper and lower bound limits the excess cost of the reconstructed probe.

\section{Optimal enclosing ellipse for a triangle}\label{app:triangle}

\begin{figure}[h]
    \centering
    \includegraphics[width=0.55\linewidth]{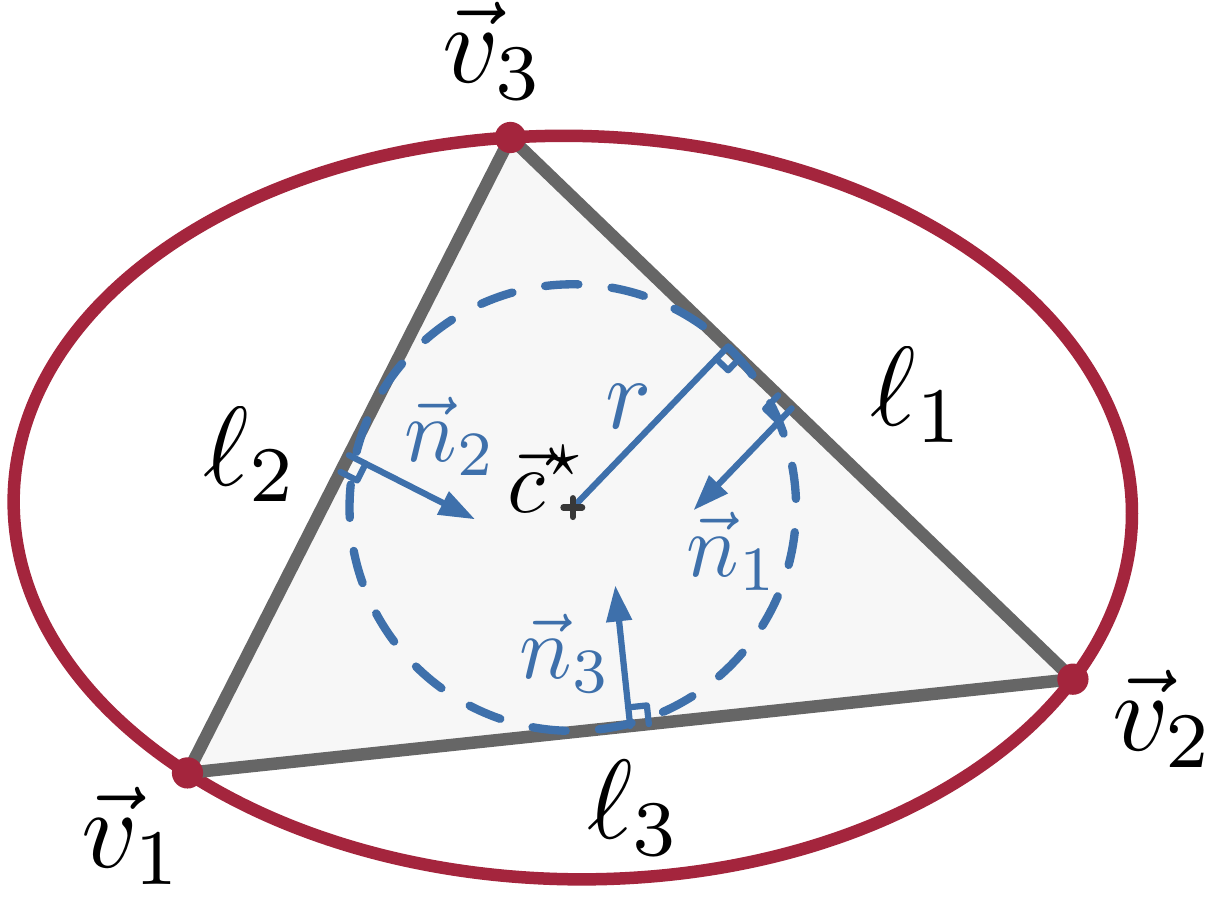}
    \caption{Triangle with vertices $\vec{\boldsymbol{v}}_i$, $i=1,2,3$.}
    \label{fig:appendix-triangle}
\end{figure}

We construct the optimal ellipse and probe when the convex hull of the joint spectrum is a triangle. Following Sec.~\ref{sec:two}, we first absorb the weight into the coordinates, $\widetilde{\vec{\boldsymbol{g}}}=W^{-1/2}\vec{\boldsymbol{g}}$. Let $\vec{\boldsymbol{v}}_i$, $i=1,2,3$, be the three vertices in these coordinates. For each vertex \(\vec{\boldsymbol{v}}_i\), let \(\ell_i\) denote the length of the side opposite \(\vec{\boldsymbol{v}}_i\), i.e., the side connecting the other two vertices, and let \(\vec{\boldsymbol{n}}_i\) denote the inward unit normal to that side (see Fig.~\ref{fig:appendix-triangle}).

The triangle has area \(A\), perimeter \(L = \sum_i \ell_i\), and inradius
\begin{equation}
r = \frac{2A}{L}.
\end{equation}
Recall that the inradius is the radius of the unique circle inside the triangle and tangent to all three sides.

An arbitrary enclosing ellipse can be written in the form
\begin{equation}
	\mathcal E=\left\{\vec{\boldsymbol{x}}:\left\|B\left(\vec{\boldsymbol{x}}-\widetilde{\vec{\boldsymbol{c}}}\right)\right\|\leq1\right\},
	\label{eq:triangle_ellipse}
\end{equation}
where $B$ is real symmetric and positive definite. This is equivalent to 
$\left(\vec{\boldsymbol{x}}-\widetilde{\vec{\boldsymbol{c}}}\right)^TS\left(\vec{\boldsymbol{x}}-\widetilde{\vec{\boldsymbol{c}}}\right)\leq 1$ with $B=\sqrt{S}$.
Its semiaxes are \(a = 1/\lambda_1\) and \(b = 1/\lambda_2\), where \(\lambda_1,\lambda_2\) are the eigenvalues of \(B\), so its harmonic mean is
\begin{equation}
h = \frac{2ab}{a+b} = \frac{2}{\operatorname{Tr}B}.
\label{eq:triangle_harmonic}
\end{equation} Minimizing the harmonic mean therefore amounts to maximizing $\operatorname{Tr}B$ while keeping all three vertices inside the ellipse.

The side lengths and inward normals of a triangle satisfy two key relations.
\begin{eqnarray}
\sum_{i=1}^{3} \ell_i \vec{\boldsymbol{n}}_i = 0,\label{eq:normals_sum}\\
\sum_{i=1}^{3} \ell_i \left(\vec{\boldsymbol{v}}_i - \widetilde{\vec{\boldsymbol{c}}}\right) \vec{\boldsymbol{n}}_i^{\mathsf T} = 2A\,I_2.
\label{eq:triangle_identity}
\end{eqnarray}
The first holds because the directed sides close the triangle. 

The second relation is a matrix identity that holds for any choice of center \(\widetilde{\vec{\boldsymbol{c}}}\). To derive it, we proceed in two steps. First, we reduce the matrix identity to an identity on a basis of \(\mathbb{R}^2\). The left-hand side of Eq.~\eqref{eq:triangle_identity} is a \(2\times 2\) matrix, so to prove that it equals \(2A\,I_2\) it suffices to show that it acts as \(2A\) times the identity on two linearly independent vectors. A convenient choice is the pair of side vectors
\begin{equation}
\vc{v}_1 - \vc{v}_2,
\qquad
\vc{v}_1 - \vc{v}_3,
\end{equation}
which span \(\mathbb{R}^2\) because the triangle is non-degenerate.

Second, we evaluate the action of the left-hand side of Eq.~\eqref{eq:triangle_identity} on an arbitrary difference \(\vc{v}_j - \vc{v}_k\):
\begin{equation}
\sum_{i=1}^{3} \ell_i \left(\vc{v}_i - \widetilde{\vc{c}}\right) \vc{n}_i^{\mathsf T}\left(\vc{v}_j - \vc{v}_k\right).
\label{eq:action-on-difference}
\end{equation}
The scalar factor in each term is \(\vc{n}_i^{\mathsf T}(\vc{v}_j - \vc{v}_k)\). Since \(\vc{n}_i\) is the unit normal to side \(i\), this inner product has a simple geometric meaning. If \(j = k\), the vector \(\vc{v}_j - \vc{v}_k\) is zero and the inner product vanishes. If \(j = i\), the inner product is the altitude from vertex \(\vc{v}_i\) to side \(i\), which equals \(2A/\ell_i\). If \(k = i\), the inner product is the negative of that altitude, \(-2A/\ell_i\). If neither \(j\) nor \(k\) equals \(i\), then \(\vc{v}_j - \vc{v}_k\) lies along side \(i\), so the inner product vanishes. Combining these cases,
\begin{equation}
\vc{n}_i^{\mathsf T}\left(\vc{v}_j - \vc{v}_k\right)
= \frac{2A}{\ell_i}\left(\delta_{ij} - \delta_{ik}\right).
\label{eq:normal-difference}
\end{equation}
Substituting Eq.~\eqref{eq:normal-difference} into Eq.~\eqref{eq:action-on-difference} yields
\begin{equation}
\begin{aligned}
&\sum_{i=1}^{3} \ell_i \left(\vc{v}_i - \widetilde{\vc{c}}\right) \frac{2A}{\ell_i}\left(\delta_{ij} - \delta_{ik}\right) \\
&= 2A\sum_{i=1}^{3} \left(\vc{v}_i - \widetilde{\vc{c}}\right)\left(\delta_{ij} - \delta_{ik}\right) \\
&= 2A\left(\vc{v}_j - \widetilde{\vc{c}}\right) - 2A\left(\vc{v}_k - \widetilde{\vc{c}}\right) \\
&= 2A\left(\vc{v}_j - \vc{v}_k\right).
\end{aligned}
\end{equation}
The center-dependent terms cancel because they appear with opposite signs. Thus
\begin{equation}
\sum_{i=1}^{3} \ell_i \left(\vc{v}_i - \widetilde{\vc{c}}\right) \vc{n}_i^{\mathsf T}\left(\vc{v}_j - \vc{v}_k\right)
= 2A\left(\vc{v}_j - \vc{v}_k\right).
\label{eq:identity-on-difference}
\end{equation}
This holds for every pair of vertices. In particular, taking \((j,k) = (1,2)\) and \((j,k) = (1,3)\) shows that $\sum_{i=1}^{3} \ell_i \left(\vec{\boldsymbol{v}}_i - \widetilde{\vec{\boldsymbol{c}}}\right) \vec{\boldsymbol{n}}_i^{\mathsf T}$ acts as \(2AI_2\) on the two linearly independent vectors \(\vc{v}_1 - \vc{v}_2\) and \(\vc{v}_1 - \vc{v}_3\). Hence
\begin{equation}
\sum_{i=1}^{3} \ell_i \left(\vc{v}_i - \widetilde{\vc{c}}\right) \vc{n}_i^{\mathsf T} = 2A\,I_2.
\end{equation}

Multiplying Eq.~\eqref{eq:triangle_identity} by $B$ and taking the trace gives
\begin{equation}
	\begin{aligned}
		2A\operatorname{Tr}B
		&=\sum_i\ell_i\vec{\boldsymbol{n}}_i^TB\left(\vec{\boldsymbol{v}}_i-\widetilde{\vec{\boldsymbol{c}}}\right)\\
		&\leq\sum_i\ell_i\left\|B\left(\vec{\boldsymbol{v}}_i-\widetilde{\vec{\boldsymbol{c}}}\right)\right\|
		\leq L.
	\end{aligned}
	\label{eq:triangle_bound}
\end{equation}
Here the last inequality uses the fact that the ellipse encloses every vertex. Its harmonic mean is therefore bounded by
\begin{equation}
	h=\frac{2}{\operatorname{Tr}B}\geq\frac{4A}{L}=2r.
	\label{eq:triangle_harmonic_bound}
\end{equation}

We now construct an ellipse that reaches this bound. Place its center at the triangle's incenter,
\begin{equation}
	\widetilde{\vec{\boldsymbol{c}}}^{\star}=\frac{1}{L}\sum_i\ell_i\vec{\boldsymbol{v}}_i,
	\label{eq:triangle_center}
\end{equation}
and choose
\begin{equation}
	B^\star=\frac{1}{2A}\sum_i\ell_i\vec{\boldsymbol{n}}_i\vec{\boldsymbol{n}}_i^T.
	\label{eq:triangle_matrix}
\end{equation}
The incenter is a distance $r$ from every side. Comparing this distance with the altitude from each vertex gives
\begin{equation}
	\vec{\boldsymbol{n}}_i^T\left(\vec{\boldsymbol{v}}_j-\widetilde{\vec{\boldsymbol{c}}}^{\star}\right)=\frac{2A}{\ell_i}\delta_{ij}-r.
	\label{eq:triangle_side_distances}
\end{equation}
Substituting into Eq.~\eqref{eq:triangle_matrix} and using $\sum_i\ell_i\vec{\boldsymbol{n}}_i=0$, we obtain $B^\star(\vec{\boldsymbol{v}}_j-\widetilde{\vec{\boldsymbol{c}}}^{\star})=\vec{\boldsymbol{n}}_j$. Since \(\|\vec{\boldsymbol{n}}_j\| = 1\), each vertex lies on the boundary of the ellipse, and the ellipse contains the entire triangle by convexity. The trace of $B^\star$ is $\operatorname{Tr}B^\star=\frac{L}{2A}$, so its harmonic mean is $h^\star=\frac{4A}{L}=2r$, which saturates the lower bound in Eq.(\ref{eq:triangle_harmonic_bound}).

The equality conditions also show that this ellipse is unique. Equality in Eq.~\eqref{eq:triangle_bound} requires 
\begin{equation}\label{eq:triangle_equality_conditions}
    B(\vec{\boldsymbol{v}}_i-\widetilde{\vec{\boldsymbol{c}}})=\vec{\boldsymbol{n}}_i
\end{equation}
at every vertex. Summing with weights $\ell_i$ gives
\begin{equation}
B\sum_{i=1}^{3} \ell_i\left(\vec{\boldsymbol{v}}_i - \widetilde{\vec{\boldsymbol{c}}}\right)
= \sum_{i=1}^{3} \ell_i \vec{\boldsymbol{n}}_i = 0,
\end{equation}
where the last equality uses the closure relation \(\sum_i \ell_i \vec{\boldsymbol{n}}_i = 0\). For any non-degenerate triangle, \(B\) is invertible, so we have $\sum_{i=1}^{3} \ell_i\left(\vec{\boldsymbol{v}}_i - \widetilde{\vec{\boldsymbol{c}}}\right)=0$, which gives
\begin{equation}
\widetilde{\vec{\boldsymbol{c}}} = \frac{1}{L}\sum_{i=1}^{3} \ell_i \vec{\boldsymbol{v}}_i = \widetilde{\vec{\boldsymbol{c}}}^{\star}.
\end{equation}
The center is thus uniquely fixed to the incenter.

We next determine \(B\). Subtracting Eq.~\eqref{eq:triangle_equality_conditions} for two different vertices \(i\) and \(j\) eliminates the center,
\begin{equation}
B\left(\vec{\boldsymbol{v}}_i - \vec{\boldsymbol{v}}_j\right) = \vec{\boldsymbol{n}}_i - \vec{\boldsymbol{n}}_j.
\label{eq:triangle_B_on_differences}
\end{equation}
Choosing two independent pairs of vertices, for instance \((i,j) = (1,2)\) and \((i,j) = (1,3)\), determines the action of \(B\) on the two linearly independent vectors \(\vec{\boldsymbol{v}}_1 - \vec{\boldsymbol{v}}_2\) and \(\vec{\boldsymbol{v}}_1 - \vec{\boldsymbol{v}}_3\). Since these vectors span \(\mathbb{R}^2\), a linear map is uniquely determined by its action on them, so \(B\) is unique. This unique \(B\) coincides with the matrix \(B^{\star}\) constructed in Eq.~\eqref{eq:triangle_matrix}. Substituting the explicit form
\begin{equation}
B^{\star} = \frac{1}{2A}\sum_{i=1}^{3} \ell_i \vec{\boldsymbol{n}}_i \vec{\boldsymbol{n}}_i^{\mathsf T}
\end{equation}
into Eq.~\eqref{eq:triangle_equality_conditions} and using Eq.~\eqref{eq:triangle_side_distances}, we find
\begin{equation}
B^{\star}\left(\vec{\boldsymbol{v}}_j - \widetilde{\vec{\boldsymbol{c}}}^{\star}\right) = \vec{\boldsymbol{n}}_j,
\qquad j = 1,2,3.
\end{equation}
Hence \(B^{\star}\) satisfies the equality conditions. Consequently, the optimal ellipse is unique, and all other points of the triangle---being convex combinations of the vertices---lie strictly inside it, with equality only at the three vertices.

The optimal probe is supported only at the three vertices. Normalization and the centroid condition uniquely fix their total populations to $p_i^\star=\ell_i/L$. The covariance expression of QFIM gives
\begin{equation}
\widetilde{F_Q}^{\star} = 4\sum_{i=1}^{3} p_i^{\star}\left(\vc{v}_i - \widetilde{\vc{c}}^{\star}\right)\left(\vc{v}_i - \widetilde{\vc{c}}^{\star}\right)^{\mathsf T}.
\label{eq:triangle_covariance_raw}
\end{equation}
This is the QFIM in the transformed coordinates with $\widetilde{\vec{\boldsymbol{g}}}=W^{-1/2}\vec{\boldsymbol{g}}$, which is connected to the QFIM in the original coordinates as $\widetilde{F_Q}^{\star}=W^{-1/2}F_Q^{\star}W^{-1/2}$. 
Using $B^\star(\vec{\boldsymbol{v}}_i-\widetilde{\vec{\boldsymbol{c}}}^{\star})=\vec{\boldsymbol{n}}_i$, we obtain
\begin{equation}
\begin{aligned}
B^{\star}\widetilde{F_Q}^{\star}B^{\star}
&= 4\sum_{i=1}^{3} p_i^{\star}\, B^{\star}\left(\vc{v}_i - \widetilde{\vc{c}}^{\star}\right)\left(\vc{v}_i - \widetilde{\vc{c}}^{\star}\right)^{\mathsf T} B^{\star} \\
&= 4\sum_{i=1}^{3} \frac{\ell_i}{L}\, \vc{n}_i \vc{n}_i^{\mathsf T}=\frac{8A}{L}B^\star.
\end{aligned}
\label{eq:triangle_covariance}
\end{equation}
This gives the QFIM
\begin{equation}
	\widetilde{F_Q}^{\star}=\frac{8A}{L}B^{\star -1},
	\label{eq:triangle_qfim}
\end{equation}
and the weighted cost
\begin{equation}
	C_W^\star=\Tr[WF_Q^{\star -1}]=\operatorname{Tr}\left[\left(\widetilde{F_Q}^{\star}\right)^{-1}\right]=\frac{L^2}{16A^2}=\frac{1}{4r^2},
	\label{eq:triangle_cost}
\end{equation}
which equals $1/h_\star^2$, confirming the constructed probe attains the optimal precision.

\section{Optimal enclosing ellipse for a parallelogram}\label{app:parallelogram}
We construct the optimal ellipse and probe when the convex hull of the joint spectrum is a parallelogram. Following Sec.~\ref{sec:two}, we first absorb the weight into the coordinates, $\widetilde{\vec{\boldsymbol{g}}}=W^{-1/2}\vec{\boldsymbol{g}}$. The convex hull remains a parallelogram after this linear transformation. Without loss of generality, we can shift the origin of the coordinates to the center of the parallelogram. The four vertices can then be represented as $\pm\vec{\boldsymbol{v}}_1$ and $\pm\vec{\boldsymbol{v}}_2$, where $\vec{\boldsymbol{v}}_1$ and $\vec{\boldsymbol{v}}_2$ are linearly independent. We now solve the enclosing problem for the parallelogram in the transformed coordinates.

Let $B=(\vec{\boldsymbol{v}}_1\ \vec{\boldsymbol{v}}_2)$ be a $2\times 2$ matrix,
and $\Delta=\det B$. As $\vec{\boldsymbol{v}}_1$ and $\vec{\boldsymbol{v}}_2$ are linearly independent, we have $\Delta\neq 0$. 

An ellipse can be written as
\begin{equation}
    (\vec{x}-\vec{\bc})^TS (\vec{x}-\vec{\bc})\leq R_S^2.
\end{equation}
Without loss of generality, we can assume $R_S^2=1$. If $R_S^2\neq 1$, it can be equivalently rewritten as
\begin{equation}
    (\vec{x}-\vec{\bc})^T\tilde{S} (\vec{x}-\vec{\bc})\leq 1,
\end{equation}
with $\tilde{S}=\frac{S}{R_S^2}$. The semiaxes of the ellipse are then given by $a=\frac{1}{\sqrt{\lambda_1}}$ and $b=\frac{1}{\sqrt{\lambda_2}}$ with $\lambda_i$ as the eigenvalues of $S$. Finding the optimal enclosing ellipse that maximizes $\frac14(\frac1a+\frac1b)^2$ is equivalent to finding $S$ with the maximal $\frac{(\Tr\sqrt{S})^2}{4}$. 

For symmetrically distributed eigenvalue vectors,
the optimal center is the origin. Indeed, containment of an opposite pair, $\pm \vec{\boldsymbol{v}}_r$ by an ellipse centered at $\vec{\boldsymbol c}$ gives
\begin{eqnarray}
    (\vec{\boldsymbol{v}}_r-\vec{\bc})^TS (\vec{\boldsymbol{v}}_r-\vec{\bc})\leq 1,\\
    (-\vec{\boldsymbol{v}}_r-\vec{\bc})^TS (-\vec{\boldsymbol{v}}_r-\vec{\bc})\leq 1.
   \end{eqnarray}
Adding these two inequalities and dividing by two gives 
$\vec{\boldsymbol{v}}_r^{\mathsf T}S\vec{\boldsymbol{v}}_r+\vec{\boldsymbol c}^{\mathsf T}S\vec{\boldsymbol c}\le1$. If $\vec{\boldsymbol c}\ne0$, the centered ellipse with matrix $S/(1-\vec{\boldsymbol c}^{\mathsf T}S\vec{\boldsymbol c})$ still encloses both pairs but with smaller semiaxes, which leads to a smaller harmonic mean of the semiaxes. For a centered ellipse, the two diagonal entries of $B^{\mathsf T}SB=\begin{pmatrix}
    \vec{\boldsymbol{v}}_1^{\mathsf T}S\vec{\boldsymbol{v}}_1 & \vec{\boldsymbol{v}}_1^{\mathsf T}S\vec{\boldsymbol{v}}_2\\
    \vec{\boldsymbol{v}}_2^{\mathsf T}S\vec{\boldsymbol{v}}_1 &\vec{\boldsymbol{v}}_2^{\mathsf T}S\vec{\boldsymbol{v}}_2
 \end{pmatrix}$ are at most one. And for the optimal enclosing ellipse, both diagonal entries must equal one. If either entry is smaller than one, we can increase that entry to one by adding a positive-semidefinite matrix to $S$. This gives another enclosing ellipse with an increased $\operatorname{Tr}\sqrt S$. Hence we need only consider
\begin{equation}
	S(z)=(B^{-1})^{\mathsf T}\begin{pmatrix}1&z\\z&1\end{pmatrix}B^{-1},\qquad |z|<1,
	\label{eq:source_fourpoint_ellipse_family}
\end{equation}
and find $z$ that maximizes $\frac{(\operatorname{Tr}\sqrt{S(z)})^2}{4}$.
With a direct computation we obtain 
\begin{equation}
	\begin{aligned}
		\frac{(\operatorname{Tr}\sqrt{S(z)})^2}{4}
		&=\frac{|\vec{\boldsymbol{v}}_1|^2+|\vec{\boldsymbol{v}}_2|^2}{4\Delta^2}+\frac{-z\vec{\boldsymbol{v}}_1^{\mathsf T}\vec{\boldsymbol{v}}_2+|\Delta|\sqrt{1-z^2}}{2\Delta^2}.
	\end{aligned}
	\label{eq:source_fourpoint_bound}
\end{equation}
Let $\vec{\boldsymbol{v}}_1^{\mathsf T}\vec{\boldsymbol{v}}_2=|\vec{\boldsymbol{v}}_1|\,|\vec{\boldsymbol{v}}_2|\cos\theta_1$,  $|\Delta|=\sqrt{|\vec{\boldsymbol{v}}_1|^2|\vec{\boldsymbol{v}}_2|^2-(\vec{\boldsymbol{v}}_1^{\mathsf T}\vec{\boldsymbol{v}}_2)^2}=|\vec{\boldsymbol{v}}_1|\,|\vec{\boldsymbol{v}}_2|\sin\theta_1$, $z=\cos\theta_2$, $\sqrt{1-z^2}=\sin\theta_2$,  here $\theta_1,\theta_2\in [0,\pi]$, then
\begin{eqnarray}
\aligned
	-z\vec{\boldsymbol{v}}_1^{\mathsf T}\vec{\boldsymbol{v}}_2+|\Delta|\sqrt{1-z^2}&=-\cos(\theta_1+\theta_2)|\vec{\boldsymbol{v}}_1|\,|\vec{\boldsymbol{v}}_2|\\
    &\le|\vec{\boldsymbol{v}}_1|\,|\vec{\boldsymbol{v}}_2|,
    \endaligned
	\label{eq:source_fourpoint_maximum}
\end{eqnarray}
the maximal value is then achieved at $\theta_2=\pi-\theta_1$, i.e., $z^\star=-\cos\theta_1=-\frac{\vec{\boldsymbol{v}}_1^{\mathsf T}\vec{\boldsymbol{v}}_2}{|\vec{\boldsymbol{v}}_1|\,|\vec{\boldsymbol{v}}_2|}$. The optimal $S$ is then given by 
\begin{equation}
	S^\star=(B^{-1})^{\mathsf T}\begin{pmatrix}1& z^\star\\z^\star&1\end{pmatrix}B^{-1}
    \label{eq:source_optimalS}
\end{equation}
with    $C_W^\star=\frac{(\Tr\sqrt{S^\star})^2}{4}=\frac{(|\vec{\boldsymbol{v}}_1|+|\vec{\boldsymbol{v}}_2|)^2}{4\Delta^2}$, here $(|\vec{\boldsymbol{v}}_1|+|\vec{\boldsymbol{v}}_2|)^2$ is the square of the sum of the distances from the center to two adjacent vertices,
4$\Delta^2$ is the square of the area of the parallelogram.

In the transformed coordinates,  $F_Q^\star=\frac{4R_{S^\star}^2}{\Tr\sqrt{S^\star}} S^{\star -\frac12}$, where $S^{\star -\frac12}$ can be computed directly as 
\begin{equation}
	(S^\star)^{-1/2}=\frac{|\vec{\boldsymbol{v}}_2|\vec{\boldsymbol{v}}_1\vec{\boldsymbol{v}}_1^{\mathsf T}+|\vec{\boldsymbol{v}}_1|\vec{\boldsymbol{v}}_2\vec{\boldsymbol{v}}_2^{\mathsf T}}{|\Delta|}.
	\label{eq:source_inverse_sqrt}
\end{equation}
Together with $\operatorname{Tr}\sqrt{S^\star}=(|\vec{\boldsymbol{v}}_1|+|\vec{\boldsymbol{v}}_2|)/|\Delta|$ and $R_{S^\star}=1$, we then have \begin{equation}
    F_Q^\star=4\frac{|\vec{\boldsymbol{v}}_2|\vec{\boldsymbol{v}}_1\vec{\boldsymbol{v}}_1^{\mathsf T}+|\vec{\boldsymbol{v}}_1|\vec{\boldsymbol{v}}_2\vec{\boldsymbol{v}}_2^{\mathsf T}}{|\vec{\boldsymbol{v}}_1|+|\vec{\boldsymbol{v}}_2|}.
\end{equation} 
Let $|\vec{\boldsymbol{v}}_j\rangle$ denote the eigenstate associated with $\vec{\boldsymbol{v}}_j$. The optimal probe state can then be written as
 \(   |\psi_{\rm opt}\rangle=\sqrt{p_1^\star} e^{i\phi_1}|\vec{\boldsymbol{v}}_1\rangle+\sqrt{p_{-1}^\star} e^{i\phi_{-1}}|-\vec{\boldsymbol{v}}_1\rangle+\sqrt{p_2^\star} e^{i\phi_{2}}|\vec{\boldsymbol{v}}_2\rangle+\sqrt{p_{-2}^\star} e^{i\phi_{-2}}|-\vec{\boldsymbol{v}}_2\rangle,\)
where the populations can be determined from the centroid condition and $F_Q^\star$ with $p_1^\star=p_{-1}^\star=\frac12\frac{|\vec{\boldsymbol{v}}_2|}{|\vec{\boldsymbol{v}}_1|+|\vec{\boldsymbol{v}}_2|}$ and $p_2^\star=p_{-2}^\star=\frac12\frac{|\vec{\boldsymbol{v}}_1|}{|\vec{\boldsymbol{v}}_1|+|\vec{\boldsymbol{v}}_2|}$.

\section{Localized-source sensing}\label{app:source}

This appendix completes the construction in Sec.~\ref{sec:source_application}. We first derive the local generators and identify the physical cost corresponding to $W=I_2$. We then determine the optimal ellipse from four spectral points, prove that it encloses the full joint spectrum, and obtain the optimal probe. An explicit measurement attains the resulting QFIM. Finally, we derive the node-separable benchmark and the limit reached as the array grows at fixed node spacing.

\subsection{Local encoding and physical precision}\label{app:source_encoding}

The known reference Hamiltonian commutes with the source Hamiltonian at every $(\beta,\xi)$. Adding the control $-\mathcal H_{\mathrm{src}}(\beta_0,\xi_0)$ therefore cancels the reference evolution exactly. In terms of the dimensionless deviations defined in Sec.~\ref{sec:source_application}, its local expansion is
\begin{equation}
	\begin{aligned}
		t[\mathcal H_{\mathrm{src}}(\beta,\xi)-\mathcal H_{\mathrm{src}}(\beta_0,\xi_0)]
		&=\sum_{k=1}^M\left(\frac{\varphi_\beta}{k^\alpha}+\frac{\alpha\varphi_\xi}{k^{\alpha+1}}\right)J_z^{(k)}\\
		&\quad+O(|\boldsymbol\varphi|^2).
	\end{aligned}
	\label{eq:source_local_expansion}
\end{equation}
Taking the derivatives at $\boldsymbol\varphi=0$ gives Eq.~\eqref{eq:source_generators}. 
To first order, the phase change at node $k$ is $\varphi_\beta/k^\alpha+\alpha\varphi_\xi/k^{\alpha+1}$. The two parameters contribute to these phases with a relative weight $\alpha/k$ that varies across the array, allowing changes in source strength and position to be distinguished when $M\ge2$.

To convert back to the physical parameters, set $L=\operatorname{diag}(t/d^\alpha,t\beta_0/d^{\alpha+1})$. Then $\boldsymbol\varphi=L(\beta-\beta_0,\xi-\xi_0)^{\mathsf T}$, and the covariance matrices satisfy
\begin{equation}
	\operatorname{Cov}\left[(\widehat\beta,\widehat\xi)\right]=L^{-1}\operatorname{Cov}(\widehat{\boldsymbol\varphi})(L^{-1})^\mathsf T.
	\label{eq:source_covariance_conversion}
\end{equation}
Thus the choice $W=I_2$ in the main text minimizes the physical cost
\begin{equation}
	\mathcal C_{I_2}=\frac{t^2}{d^{2\alpha}}\operatorname{Var}(\widehat\beta)+\frac{t^2\beta_0^2}{d^{2\alpha+2}}\operatorname{Var}(\widehat\xi).
	\label{eq:source_physical_cost}
\end{equation}
For a different prescribed physical weight $W_{\beta,\xi}$, the corresponding dimensionless weight is $W=(L^{-1})^\mathsf TW_{\beta,\xi}L^{-1}$.

\subsection{The optimal enclosing ellipse}\label{app:source_ellipse}

We now determine the optimal enclosing ellipse for the eigenvalue vectors,
\[
\vec{\boldsymbol{g}}_i=\sum_{k=1}^M s_k\,k^{-\alpha}
\begin{pmatrix}1\\ \alpha/k\end{pmatrix},
\qquad s_k\in\{-J,-J+1,\ldots,J\}.
\]
The extreme points of the convex hull of these eigenvalue vectors are given by
\(
\{\pm\,\vec{\boldsymbol{v}}_r|
 r=1,\ldots,M\}
\)
where
\begin{equation}
    \vec{\boldsymbol{v}}_r
    =
    \sum_{k=1}^{r}J\,k^{-\alpha}
    \begin{pmatrix}1\\ \alpha/k\end{pmatrix}
    +
    \sum_{k=r+1}^{M}(-J)\,k^{-\alpha}
    \begin{pmatrix}1\\ \alpha/k\end{pmatrix}.
    \label{eq:vertices}
\end{equation}
These are the eigenvalue vectors associated with common eigenstates
\(
|{+}J\rangle^{\otimes r}|{-}J\rangle^{\otimes (M-r)}\) and \(|{-}J\rangle^{\otimes r}|{+}J\rangle^{\otimes (M-r)}
\)
for \(r=0,1,\ldots,M\). The case \(r=0\) gives the all-down state and coincides with \(-\vec{\boldsymbol{v}}_M\), while \(r=M\) gives \(\vec{\boldsymbol{v}}_M\). Hence the distinct extreme points are the \(2M\) vectors \(\{\pm\vec{\boldsymbol{v}}_r\), \(r=1,\ldots,M\}\).

We first solve the enclosing problem for $\pm\vec{\boldsymbol{v}}_1$ and $\pm\vec{\boldsymbol{v}}_M$, which form a parallelogram. The optimal enclosing ellipse is then given by 
\begin{equation}
    \vec{v}^TS^\star \vec{v}\leq 1,
\end{equation}
where 
\begin{equation}
	S^\star=(B^{-1})^{\mathsf T}\begin{pmatrix}1&z^\star\\z^\star &1\end{pmatrix}B^{-1},
    \end{equation}
$B=(\vec{\boldsymbol{v}}_1\ \vec{\boldsymbol{v}}_M)$ and $z^\star=-\vec{\boldsymbol{v}}_1^{\mathsf T}\vec{\boldsymbol{v}}_M/(|\vec{\boldsymbol{v}}_1||\vec{\boldsymbol{v}}_M|)$.

We now show that this ellipse also contains every intermediate vertex. 
The four points $\pm\vec{\boldsymbol{v}}_1$ and $\pm\vec{\boldsymbol{v}}_M$ already lie on its boundary. To check the intermediate vertices of $1<r<M$, we use Eq.~\eqref{eq:source_optimalS} to write
\begin{equation}
	\vec{\boldsymbol{v}}_r^{\mathsf T}S^\star\vec{\boldsymbol{v}}_r=\left|B^{-1}\vec{\boldsymbol{v}}_r\right|^2-2\eta\left[B^{-1}\vec{\boldsymbol{v}}_r\right]_1\left[B^{-1}\vec{\boldsymbol{v}}_r\right]_2.
	\label{eq:source_containment_form}
\end{equation}
For $\alpha\ge2$, the bound $\sum_{k=2}^M k^{-\alpha}<1$ ensures that both components of $\vec{\boldsymbol{v}}_1$ and $\vec{\boldsymbol{v}}_M$ are positive, giving $\eta>0$. The last term therefore lowers the quadratic form whenever the two coordinates of $B^{-1}\vec{\boldsymbol{v}}_r$ are positive. To establish containment, it suffices to show that these coordinates are positive and that $\left|B^{-1}\vec{\boldsymbol{v}}_r\right|^2<1$ for $1<r<M$.

For $1<r<M$, write
\begin{equation}
    L_r=\sum_{k=2}^r\frac{k-1}{k^{\alpha+1}},\qquad R_r=\sum_{l=r+1}^M\frac{l-1}{l^{\alpha+1}}.
\end{equation}
Substituting Eq.~\eqref{eq:vertices} into $B^{-1}\vec{\boldsymbol{v}}_r$ gives
\begin{equation}
	B^{-1}\vec{\boldsymbol{v}}_r=\frac{1}{L_r+R_r}\left[\begin{pmatrix}R_r\\L_r\end{pmatrix}+\sum_{k=2}^r\sum_{l=r+1}^M\frac{l-k}{k^{\alpha+1}l^{\alpha+1}}\begin{pmatrix}1\\1\end{pmatrix}\right].
	\label{eq:source_partial_coordinates}
\end{equation}
Both coordinates are positive. Since $0<l-k<(k-1)(l-1)$ for $2\le k<l$, the summation satisfies
\begin{equation}
    0<\sum_{k=2}^r\sum_{l=r+1}^M\frac{l-k}{k^{\alpha+1}l^{\alpha+1}}<L_rR_r.
\end{equation}
Moreover,
\begin{equation}
	\begin{aligned}
		L_r+R_r&\le\sum_{k=2}^M\frac{k-1}{k^3}\\
		&<\sum_{k=2}^M\frac{1}{k(k+1)}=\frac12-\frac{1}{M+1}<\frac12.
	\end{aligned}
	\label{eq:source_tail_bound}
\end{equation}
Using the bounds above, we therefore obtain
\begin{equation}
	\begin{aligned}
		\vec{\boldsymbol{v}}_r^{\mathsf T}S^\star\vec{\boldsymbol{v}}_r
		&<\left|B^{-1}\vec{\boldsymbol{v}}_r\right|^2\\
		&<\frac{R_r^2(1+L_r)^2+L_r^2(1+R_r)^2}{(L_r+R_r)^2}\\
		&=1-\frac{2L_rR_r(1-L_r-R_r-L_rR_r)}{(L_r+R_r)^2}<1.
	\end{aligned}
	\label{eq:source_strict_containment}
\end{equation}
The last inequality follows from $L_r+R_r<1/2$ and $L_rR_r\le(L_r+R_r)^2/4$. Thus every intermediate vertex, together with its opposite, lies strictly inside the ellipse. For $M=2$, there are no intermediate vertices.

Since the ellipse is convex, enclosing all vertices also encloses the full joint spectrum. Its only contact points are $\pm\vec{\boldsymbol{v}}_1$ and $\pm\vec{\boldsymbol{v}}_M$. Every ellipse enclosing the full spectrum must also enclose these four points, so the ellipse optimized for them is optimal for the full spectrum as well.

Let $p_{r,+}$ and $p_{r,-}$ be the populations at $\pm\vec{\boldsymbol{v}}_r$, for $r=1,M$. 
The optimal probe state has 
$p_{1,+}^\star=p_{1,-}^\star=\frac12\frac{|\vec{\boldsymbol{v}}_M|}{|\vec{\boldsymbol{v}}_1|+|\vec{\boldsymbol{v}}_M|}$ and $p_{M,+}^\star=p_{M,-}^\star=\frac12\frac{|\vec{\boldsymbol{v}}_1|}{|\vec{\boldsymbol{v}}_1|+|\vec{\boldsymbol{v}}_M|}$, which gives 
\begin{equation}
    F_Q^\star=4\frac{|\vec{\boldsymbol{v}}_M|\vec{\boldsymbol{v}}_1\vec{\boldsymbol{v}}_1^{\mathsf T}+|\vec{\boldsymbol{v}}_1|\vec{\boldsymbol{v}}_M\vec{\boldsymbol{v}}_M^{\mathsf T}}{|\vec{\boldsymbol{v}}_1|+|\vec{\boldsymbol{v}}_M|}.
\end{equation}

We also construct an explicit measurement.  Write $|D_M\rangle=|+J\rangle^{\otimes M}$ and $|\overline D_M\rangle=|-J\rangle^{\otimes M}$. For local estimation at $\boldsymbol\varphi=0$, use the projectors onto
\begin{equation}
	|m_{r,\pm}\rangle=\frac{|D_r\rangle\pm i e^{i\chi_r}|\overline D_r\rangle}{\sqrt2},\qquad r=1,M.
	\label{eq:source_optimal_measurement}
\end{equation}
These four orthonormal states span the support of the probe. With the reference evolution canceled, Eq.~\eqref{eq:source_local_expansion} gives the outcome probabilities near the operating point,
\begin{equation}
	P_{r,\pm}(\boldsymbol\varphi)=\frac{p_r^\star}{2}\left[1\pm2\vec{\boldsymbol{v}}_r^{\mathsf T}\boldsymbol\varphi\right]+O(|\boldsymbol\varphi|^2).
	\label{eq:source_measurement_probabilities}
\end{equation}
The relative phase $\chi$ between the two pairs does not enter these probabilities. Substituting them into Eq.~\eqref{eq:cfi} gives
\begin{equation}
	\left.F_C\right|_{\boldsymbol\varphi=0}=4\sum_{r=1,M}p_r^\star\vec{\boldsymbol{v}}_r\vec{\boldsymbol{v}}_r^{\mathsf T}=F_Q^\star.
	\label{eq:source_measurement_cfim}
\end{equation}
The attaining covariance in physical units is therefore $L^{-1}(F_Q^\star)^{-1}L^{-\mathsf T}/\nu$ in the local asymptotic regime, and the minimum of Eq.~\eqref{eq:source_physical_cost} is $C_{I_2}^\star/\nu$.

\subsection{Node-separable precision benchmark}\label{app:source_benchmark}

We derive Eq.~\eqref{eq:source_separable_cost} allowing arbitrary states separable between nodes, $\rho_{\mathrm{sep}}=\sum_\ell q_\ell\bigotimes_{k=1}^M\rho_{k,\ell}$, with $q_\ell\ge0$ and $\sum_\ell q_\ell=1$. To first order at the operating point, both parameters enter node $k$ through the local phase $\vec{\boldsymbol r}_k^{\mathsf T}\boldsymbol\varphi$. If $f_{k,\ell}$ is the QFI of $\rho_{k,\ell}$ for this phase, additivity gives
\begin{equation}
	F_Q\left(\bigotimes_{k=1}^M\rho_{k,\ell}\right)=\sum_{k=1}^M f_{k,\ell}\vec{\boldsymbol r}_k\vec{\boldsymbol r}_k^{\mathsf T}.
	\label{eq:source_product_qfim}
\end{equation}
The extremal eigenvalues $\pm J$ of $J_z^{(k)}$ imply $f_{k,\ell}\le4\operatorname{Var}_{\rho_{k,\ell}}(J_z^{(k)})\le4J^2$. Convexity of the QFIM~\cite{LiuYuanLuWang2020} then yields
\begin{equation}
	\begin{aligned}
		F_Q(\rho_{\mathrm{sep}})&\preceq\sum_\ell q_\ell F_Q\left(\bigotimes_{k=1}^M\rho_{k,\ell}\right)\\
		&\preceq4J^2\sum_{k=1}^M\vec{\boldsymbol r}_k\vec{\boldsymbol r}_k^{\mathsf T}.
	\end{aligned}
	\label{eq:source_separable_qfim}
\end{equation}
The product state given above Eq.~\eqref{eq:source_separable_cost} attains this matrix bound because every local superposition has $f_k=4J^2$. Its QFIM is nonsingular for $M\ge2$, since the response vectors are not all parallel, and the commuting generators ensure local attainability. Since matrix inversion reverses the positive-definite order, Eq.~\eqref{eq:source_separable_qfim} proves Eq.~\eqref{eq:source_separable_cost}. In particular, classical correlations between nodes cannot lower this benchmark.

\subsection{Precision at fixed node spacing}\label{app:source_scaling}

We take $M\to\infty$ with $d$ and $J$ fixed in the array $x_k-\xi_0=kd$. For any $s>1$, the response tail satisfies
\begin{equation}
	\sum_{k=M+1}^{\infty}k^{-s}\le\int_M^\infty x^{-s}\,dx=\frac{M^{1-s}}{s-1}.
	\label{eq:source_response_tails}
\end{equation}
For $\alpha\ge2$, taking $s=\alpha$ and $s=\alpha+1$ shows that $\vec{\boldsymbol v}_M$ in Eq.~\eqref{eq:source_aligned_vertex} and $\vec{\boldsymbol v}_1=2J(1,\alpha)^{\mathsf T}-\vec{\boldsymbol v}_M$ approach finite limits, with corrections of order $M^{1-\alpha}$. Their determinant remains nonzero in this limit because
\begin{equation}
	|\det(\vec{\boldsymbol v}_1,\vec{\boldsymbol v}_M)|=2J^2\alpha\sum_{k=2}^M\frac{k-1}{k^{\alpha+1}}\ge\frac{J^2\alpha}{2^\alpha}>0.
	\label{eq:source_limiting_determinant}
\end{equation}
The cost in Eq.~\eqref{eq:source_precision_lower_bound} is therefore a smooth function of the two limiting contact vectors, which proves Eq.~\eqref{eq:source_fixed_spacing_limit}. The value of $C_{I_2,\infty}^\star$ is obtained by replacing the finite sums in the two contact vectors by their infinite sums in Eq.~\eqref{eq:source_precision_lower_bound}.

\bibliography{reference}

\end{document}